\documentclass[iop,apj]{emulateapj}

\usepackage{mathptmx}
\usepackage{wasysym} 
\usepackage{amssymb}
\usepackage{color}
\usepackage{subfigure}

\journalinfo{Draft version \today}
\slugcomment{Accepted for publication in ApJ}
\shorttitle{X-RAY PROPERTIES OF SDSS MINI-BAL QUASARS}
\shortauthors{WU ET AL.}

\begin{document}


\title{The X-Ray Properties of the Optically Brightest Mini-BAL Quasars
       \\ from the Sloan Digital Sky Survey}


\author{
Jianfeng~Wu\altaffilmark{1,2}, 
W.~N.~Brandt\altaffilmark{1,2},
M.~L.~Comins\altaffilmark{1},
Robert~R.~Gibson\altaffilmark{3},
Ohad~Shemmer\altaffilmark{4},
Gordon~P.~Garmire\altaffilmark{1},
Donald~P.~Schneider\altaffilmark{1}
}

\altaffiltext{1}
    {Department of Astronomy \& Astrophysics, The Pennsylvania State
     University, 525 Davey Lab, University Park, PA 16802, USA}
\altaffiltext{2}
    {Institute for Gravitation and the Cosmos, The Pennsylvania State 
     University, University Park, PA 16802, USA}
\altaffiltext{3}
    {Department of Astronomy, University of Washington, Box 351580, 
     Seattle, WA 98195, USA}
\altaffiltext{4}
    {Department of Physics, University of North Texas, Denton, TX 76203, USA}


\begin{abstract}
\noindent We have compiled a sample of 14 of the optically brightest 
radio-quiet quasars ($m_{i}$~$\le$~17.5 and $z$~$\ge$~1.9) in the Sloan Digital 
Sky Survey Data Release 5 quasar catalog that have \ion{C}{4} mini-BALs present
in their spectra. \hbox{X-ray} data for 12 of the objects were 
obtained via a {\it Chandra} snapshot survey using ACIS-S, while data 
for the other two quasars were obtained from archival {\it XMM-Newton} 
observations. Joint \hbox{X-ray} spectral analysis shows the mini-BAL quasars have
a similar average power-law photon index ($\Gamma\approx1.9$) and level of 
intrinsic absorption
($N_H \lesssim 8\times 10^{21} \ {\rm cm}^{-2}$) as \hbox{non-BMB} 
(neither BAL nor mini-BAL) quasars. 
Mini-BAL quasars are more similar to non-BMB quasars than to BAL quasars in
their distribution of relative \hbox{X-ray} brightness (assessed with
 $\Delta\alpha_{\rm ox}$). Relative colors indicate mild dust reddening in the optical
spectra of mini-BAL quasars. 
Significant correlations between $\Delta\alpha_{\rm ox}$ and UV absorption properties
are confirmed for a sample of 56 sources combining mini-BAL and BAL quasars with high
signal-to-noise ratio rest-frame UV spectra, which generally supports models in which \hbox{X-ray} 
absorption is important in enabling driving of the UV absorption-line wind. 
We also propose alternative parametrizations of the UV 
absorption properties of mini-BAL and BAL quasars, which may better
describe the broad absorption troughs in some respects. 
\end{abstract}

\keywords{galaxies: active --- galaxies: nuclei --- quasars: absorption lines ---
quasars: emission lines --- \hbox{X-ray}s: galaxies}


\section{Introduction}
Absorption lines in quasar spectra provide useful probes of outflows in 
active galactic nuclei (AGNs). Broad absorption lines (BALs) are traditionally 
defined to have velocity widths $\ge$~2,000~km~s$^{-1}$ (e.g., Weymann et al. 
1991) in transitions such as \ion{C}{4}~$\lambda$1549, 
\ion{Si}{4}~$\lambda$1400, \ion{Al}{3}~$\lambda$1857, and 
\ion{Mg}{2}~$\lambda$2799. The number of quasars with intermediate velocity width absorption 
lines (1,000$-$2,000~km~s$^{-1}$), known as mini-BALs, is comparable to or 
greater than the number with traditional BALs (e.g., Trump et al. 
2006; Gibson et al. 2009a, hereafter G09). BAL and mini-BAL outflows are powerful in luminous 
quasars, partly indicated by their high velocities, and they could be 
responsible for significant feedback into quasar host galaxies 
(e.g., Ganguly \& Brotherton 2008; Brandt et al. 2009). 
These outflows are often envisioned as radiatively accelerated equatorial disk 
winds, and it appears that they must be significantly shielded from the 
soft \hbox{X-ray} emission from the central region in order to avoid 
overionization of the outflow, which would effectively prevent radiative 
acceleration (e.g., Murray et al. 1995; Proga et al. 2000).
Multiwavelength observations of both 
BAL and mini-BAL quasars must be compiled in order to determine the 
physical characteristics and processes governing AGN outflows. Furthermore, 
multiwavelength data will also help determine whether 
mini-BALs are part of a continuum of absorption phenomena whose progenitor is 
the same as BALs and narrow absorption lines (NALs), or if mini-BALs instead 
constitute a different physical phenomenon.

It has been suggested (e.g., Knigge et al. 2008) that mini-BAL quasars are 
part of a population that is different from BAL quasars, but UV absorption-line 
strength and maximum outflow velocity appear to be correlated with relative 
\hbox{X-ray} brightness (i.e., an estimate of the level of \hbox{X-ray} absorption), 
with mini-BAL quasars 
generally being intermediate between BAL and \hbox{non-BMB} 
(meaning neither BAL nor mini-BAL) quasars (e.g., Gallagher 
et al. 2006; G09). The effective hard \hbox{X-ray} photon index, $\Gamma$, is also 
correlated with the relative \hbox{X-ray} brightness for combined radio-quiet BAL, 
mini-BAL, and non-BMB quasar samples. These results suggest a physical link between BALs 
and mini-BALs. On the other hand, there are also cases in which the outflow velocity 
for a mini-BAL is higher than expected for a given \hbox{X-ray} brightness, which 
indicates that some mini-BALs may not require strong \hbox{X-ray} absorption or may be 
launched in a different way than is typical (G09). Furthermore, the classical 
definition of ``mini-BAL'' using absorption index ($AI$, see \S 2) cannot fully address
the complex absorption features in quasar spectra (e.g., G09).

In this paper, we study the \hbox{X-ray} properties of 14 of the optically brightest 
radio-quiet quasars (RQQs) whose spectra contain a \ion{C}{4} mini-BAL
in the Sloan Digital Sky Survey (SDSS; York et~al. 2000) Data Release 5 (DR5) quasar catalog 
(Schneider et al. 2007) using a combination of new {\it Chandra} snapshot 
observations and archival {\it XMM-Newton} observations. We seek to extend the 
\hbox{X-ray} coverage of mini-BAL quasars, to study nuclear outflows, 
and to evaluate the properties of mini-BAL quasars relative to 
BAL and non-BMB quasars. Compared to the mini-BAL quasars in G09, our 
more luminous sources cover a different portion of the luminosity-redshift plane (see Fig.~1$a$)
which has not been explored. Our sources have uniform, high-quality X-ray data and a 100\%
detection rate (see \S 3). Furthermore, our sources are among the optically brightest \hbox{mini-BAL} quasars
(see Fig.~1$b$), which are also generally likely to be the X-ray brightest. They may be the best objects
for study in the X-ray band using future missions, e.g., the International X-ray Observatory 
({\it IXO}; e.g., White et~al.~2010), as well as in the optical band.

Unless otherwise specified, a cosmology with 
$H_{0}$~=~70~km~s$^{-1}$~Mpc$^{-1}$, $\Omega_{\rm M}\;=\;0.3$, and 
$\Omega_{\Lambda}\;=\;0.7$ will be assumed. Furthermore, the term ``HiBAL'' 
refers to BAL quasars with high-ionization states only (e.g., \ion{C}{4}), 
while ``LoBAL'' refers to BAL quasars with low-ionization states (e.g., 
\ion{Al}{3} or \ion{Mg}{2}) and usually also high-ionization states (e.g., Gibson et~al.~2009b).
We define positive velocities for material flowing outward with respect to the quasar 
emission rest frame. \S 2 discusses 
our sample selection; \S 3 details our \hbox{X-ray} data reduction
and analysis; and \S 4 discusses the implications of our results. \S 5 is a 
summary of our conclusions. 


\section{Sample Selection}
We sorted the SDSS DR5 quasar catalog (Schneider et al. 2007), which 
covers $\approx5740$ deg$^2$, by $m_{i}$ 
and identified all of the mini-BAL quasars with $m_{i}\;\le$ 17.5 and $z\;\ge\;1.9$,
which resulted in a list of 29 robustly identified quasars whose mini-BAL 
nature (as described further below) is insensitive to reasonable continuum placement. The redshift 
requirement ensures that all objects considered have high-quality \ion{C}{4} 
coverage in their SDSS spectra. The magnitude limit was chosen to provide a 
sample size that could be observed with a practical amount of {\it Chandra} 
observation time. The SDSS quasar catalog BEST photometric apparent $i$-band 
magnitudes were used wherever possible, but if the BEST $m_{i}\;=0$, then the 
TARGET value of $m_{i}$ was used. Also, we sought radio-quiet objects that 
were not detected by the {\it FIRST} survey (Becker et al. 1995). ``Radio-quiet'' 
is commonly defined as $R < 10$, where $R$ is calculated by Equation (5) in \S 3. 
We required our potential targets to be covered, but undetected by the {\it FIRST} 
survey to 
minimize any \hbox{X-ray} emission contribution from possible jets associated with 
the quasars. Six of the 29
objects had {\it FIRST} detections and were therefore excluded. 
Furthermore, two of the radio-quiet mini-BAL quasars on our list already had 
\hbox{X-ray} coverage; we included these archival objects in our sample but did not 
propose new targeted {\it Chandra} observations for them. We 
proposed short ``snapshot'' (4$-$6 ks) {\it Chandra} observations of the 
resulting 21 mini-BAL quasars, generally prioritizing the targets by their 
brightness ($m_{i}$). We were awarded observing time in Cycle 10 for 12 of 
the brightest objects on the target list. 

Our resulting sample thus consists of 14 of the optically brightest mini-BAL 
quasars in the SDSS DR5 quasar catalog, as shown in 
Fig.~1 (12 new {\it Chandra} observations and two archival {\it XMM-Newton}
observations). The {\it XMM-Newton} data for SDSS J$142656.17+602550.8$, also known as 
SBS $1425+606$, were previously published in Shemmer et al.~(2008). Relevant 
\hbox{X-ray} information for SDSS J$160222.72+084538.4$, observed by {\it XMM-Newton} at 
17.0$'$ off-axis, is located in the {\it XMM-Newton} Serendipitous Source 
Catalog (2XMMi; Watson et al. 2009). For both of the {\it XMM-Newton} 
observations, only data from the \verb+pn+ detector were used, because of its high 
sensitivity and number of counts with respect to the \verb+MOS+ detectors. An 
observing log is shown in Table~1. Compared to the mini-BAL quasar sample 
(48 objects, all of which have X-ray coverage) utilized in G09, nearly all of the objects in our 
sample are brighter in both apparent $i$-band magnitude, $m_{i}$, and 
absolute $i$-band magnitude, $M_{i}$. There are, however, two objects from 
G09 in Fig.~1{\it (b)} that are comparably bright in $m_{i}$ to our 
sample objects. One is the well-known gravitationally lensed quasar 
PG~1115+080 (e.g., Michalitsianos et al. 
1996; Chartas et al. 2007). This object was not included in our sample because 
its redshift of $1.74$ does not meet our 
$z\;\ge\;1.9$ criterion. The other object, SDSS~J$231324.45+003444.5$, likely has 
a high-velocity \ion{C}{4}~BAL in its spectrum, but was included in the sample 
from G09 as an ``ambiguous'' mini-BAL source. See \S 
2.1 in G09 for a more detailed discussion regarding this source.
\begin{deluxetable}{llccc}
\tablecolumns{5} \tablenum{1} \tabletypesize{\footnotesize}
\tablewidth{0pt}
\tablecaption{Observation Log}
\tablehead{
    \colhead{} & \colhead{} & \colhead{} 
               & \colhead{Observation} & \colhead{Exp. Time} \\
    \colhead{} & \colhead{Object Name (SDSS J)} & \colhead{$z$} 
               & \colhead{Date}  & \colhead{(ks)\tablenotemark{a}}
}
\startdata
\multicolumn{2}{l}{{\it Chandra} Cycle 10 Objects} & & & \\
& $080117.79+521034.5$ & $3.23$ & 2009 Jan 12 & $4.1$ \\	
& $091342.48+372603.3$ & $2.13$ & 2009 Jan 22 & $5.1$ \\
& $092914.49+282529.1$ & $3.41$ & 2009 Jan 07 & $5.0$ \\
& $093207.46+365745.5$ & $2.90$ & 2009 Jan 09 & $5.5$ \\
& $105158.74+401736.7$ & $2.17$ & 2009 Jan 20 & $4.1$ \\
& $105904.68+121024.0$ & $2.50$ & 2009 Feb 01 & $4.1$ \\
& $120331.29+152254.7$ & $2.99$ & 2009 Mar 14 & $4.1$ \\
& $123011.84+401442.9$ & $2.05$ & 2009 Oct 12 & $5.0$ \\
& $125230.84+142609.2$ & $1.94$ & 2009 Feb 28 & $4.0$ \\
& $141028.14+135950.2$ & $2.22$ & 2009 Nov 28 & $4.1$ \\	
& $151451.77+311654.0$ & $2.14$ & 2008 Dec 06 & $4.1$ \\
& $224649.29-004954.3$ & $2.04$ & 2009 Jan 02 & $5.1$ \\ \\
\multicolumn{2}{l}{Archival {\it XMM-Newton} Objects} & & & \\
& $142656.17+602550.8$\tablenotemark{b} & $3.20$ & 2006 Nov 12 & $7.0$ \\
& $160222.72+084538.4$ & $2.28$ & 2003 Aug 09 & $13.3$
\enddata
\tablenotetext{a}{The {\it Chandra} and {\it XMM-Newton} exposure times are 
corrected for detector dead time.}
\tablenotetext{b}{The data for SDSS J$142656.17+602550.8$ were originally 
published in Shemmer et al.~(2008), and it is also known as SBS $1425+606$. 
The exposure time used here is the pn detector exposure listed in Table 1 in 
Shemmer et al.~(2008); exposure times for the MOS1 and the MOS2 detectors were 
20.6 ks and 20.9 ks, respectively.}
\end{deluxetable}\label{table1}
\begin{deluxetable*}{clrrrrcl}
\tablecolumns{8} \tablenum{2} \tabletypesize{\scriptsize}
\tablewidth{0pt}
\tablecaption{\ion{C}{4} Mini-BAL Properties}
\tablehead{
    \colhead{} & \colhead{}                     & \colhead{$AI$}         
    & \colhead{$v_{\rm min}$} & \colhead{$v_{\rm max}$}  & \colhead{$v_{\rm wt}$\tablenotemark{a}}
    & \colhead{No. of Troughs} & \colhead{}\\ 
    \colhead{} & \colhead{Object Name (SDSS J)} & \colhead{(km~s$^{-1}$)} 
    & \colhead{(km~s$^{-1}$)} & \colhead{(km~s$^{-1}$)} 
    & \colhead{(km~s$^{-1}$)}
    & \colhead{Contributing to $AI$} & \colhead{Notes\tablenotemark{b}}
}
\startdata
\multicolumn{2}{l}{{\it Chandra} Cycle 10 Objects} & & & \\
& $080117.79+521034.5$ & $251.0$  & $26036  $ & $27132$ & $26641$ & 1 & \\
& $091342.48+372603.3$ & $625.8$  & $10767  $ & $23014$ & $14463$ & 2 & Feature at 
                                                                $\approx$1535~
                                                                \AA~ fails 
                                                                width 
                                                                criterion\\
& $092914.49+282529.1$ & $332.2$  & $23881  $ & $25047$ & $24483$ & 1 & \\
& $093207.46+365745.5$ & $533.6$  & $3146$  & $4596$    & $3871$ & 1 & Features at 
                                                                $\approx$1480~
                                                                \AA~ and 
                                                                $\approx$1510~
                                                                \AA~ \\
&                      &          &         &           &        &    & fail width
                                                                and depth 
                                                                criteria\\
& $105158.74+401736.7$ & $239.3$  & $1805$  & $2910$    & $2383$ & 1 & \\
& $105904.68+121024.0$ & $2384.7$ & $5378$  & $23349$   & $11445$ & 5 & \\
& $120331.29+152254.7$ & $311.3$  & $21530$ & $22629$   & $22118$ & 1 & \\
& $123011.84+401442.9$ & $556.8$  & $1710$  & $2953$    & $2326$ & 1 & \\
& $125230.84+142609.2$ & $528.7$  & $2165$  & $3615$    & $2914$ & 1 & Features at 
                                                                $\approx$1440~
                                                                \AA~ and 
                                                                $\approx$1520~
                                                                \AA~ \\
&                      &          &         &           &       &    & fail depth
                                                                and width 
                                                                criteria\\
& $141028.14+135950.2$ & $191.6$  & $1842$   & $2946$   & $2394$ & 1 & \\
& $151451.77+311654.0$ & $456.0$  & $25290$ & $27003$   & $26188$ & 1 & Feature at 
                                                              $\approx$1430~
                                                              \AA~ fails both depth 
                                                              and \\ 
&                      &          &         &           &        &    & width 
                                                              criteria\\
& $224649.29-004954.3$ & $668.5$  & $4199$  & $5648$    & $4858$ & 1 & \\ \\
\multicolumn{2}{l}{Archival {\it XMM-Newton} Objects} & & & \\
& $142656.17+602550.8$ & $356.0$  & $27229$ & $28324$   & $27811$ & 1 & Feature at 
                                                                $\approx$1525~
                                                                \AA~ fails 
                                                                width 
                                                                criterion\\
& $160222.72+084538.4$ & $316.4$  & $14068$ & $15239$   & $14689$ & 1 & Feature at 
                                                                $\approx$1540~
                                                                \AA~ fails 
                                                                width 
                                                                criterion
\enddata
\tablenotetext{a}{The weighted average velocity of mini-BAL troughs. See specific
definition in \S4.4.}
\tablenotetext{b}{This column contains explanations as to why some 
non-negligible absorption troughs seen in Fig.~2 were not included in the 
absorption index, $AI$. A trough is excluded from $AI$ if it fails the width 
criterion and/or the depth criterion. The ``width criterion'' requires that 
the trough have a velocity width between 1,000$-$2,000~km~s$^{-1}$. The ``depth 
criterion'' requires that the trough remain at least 10\% below the continuum. 
The depth criterion can be failed if there is a peak inside the absorption 
trough that rises above this level.}
\end{deluxetable*}
The two quantities that were used in distinguishing mini-BALs from BALs and 
non-BMBs are the extended balnicity index ($BI_{0}$) and the absorption index 
($AI$). $BI_{0}$ is defined as
\begin{equation}
    BI_{0} = \int_{0}^{25,000} \left(1 - \frac{f(v)}{0.9}\right)C\ dv,
\end{equation} 
where $f(v)$ is the ratio of the observed spectrum to the continuum model for 
the spectrum as a function of velocity. The continuum model fit to the SDSS 
spectra was determined using the algorithm described in \S 2.1 of Gibson 
et al.~(2008a). $C$ is a parameter used to indicate BAL absorption. $C$~=~1 if 
the spectrum is at least 10\% below the continuum model for velocity widths of 
at least 2,000~km~s$^{-1}$; $C$~=~0 otherwise. The integration limits here are 
the same as in G09, which differ from those used in Weymann 
et al.~(1991), but allow characterization of BALs even at low outflow 
velocities.

$AI$ is defined as
\begin{equation}
    AI = \int_{0}^{29,000} (1 - f(v))\ C'dv,
\end{equation} 
where $C'$~=~1 when the velocity width is at least 1,000~km~s$^{-1}$
and the absorption trough falls at least 10\% below the continuum; $C'$~=~0 
otherwise. The integration limits for $AI$ are chosen such that the range of 
\ion{C}{4} outflow velocities is maximized and does not include the 
\ion{Si}{4} $\lambda$1400 emission line (Trump et al. 2006). BALs have 
$AI\;>\;0$ and $BI_{0}\;>\;0$; mini-BALs have $AI\;>\;0$ and $BI_{0}\;=\;0$; 
and non-BMBs have $AI\;=\;0$ and $BI_{0}\;=\;0$. Thus, the objects in our 
sample were selected such that, for \ion{C}{4} $\lambda$1549, $BI_{0}\;=\;0$ 
and $AI\;>\;0$. All measurements of $BI_{0}$, $AI$, and other absorption 
parameters are made in the rest frame of the quasar as defined by the improved
redshift measurement of Hewett \& Wild (2010). One mini-BAL quasar in G09, 
SDSS~J142301.08+533311.8, is classified as a \hbox{non-BMB} quasar using this improved
redshift value because its only broad 
absorption trough lies above the velocity integration limit of $29,000$~km~s$^{-1}$ 
in the $AI$ definition.

The SDSS spectra for our sample are shown in Fig.~2. These spectra are 
corrected for Galactic extinction according to Cardelli et al.~(1989) and for 
fiber light loss. The light-loss correction was calculated as the average 
difference between the synthetic and photometric $g$, $r$, and $i$ magnitudes 
for each quasar, as described in \S 3 of Just et al.~(2007). The 
synthetic magnitudes are provided by the SDSS data pipeline and correspond to 
the integrated flux over the $g$, $r$, and $i$ bandpasses in the SDSS spectrum. 
It is assumed that there is no variation in the flux between the SDSS 
photometric and spectroscopic epochs.

The rest wavelength of \ion{C}{4} $\lambda$1549 is indicated in Fig.~2, as 
well as the minimum ($v_{\rm min}$) and maximum ($v_{\rm max}$) outflow 
velocities. The wavelengths of absorption troughs can be converted to outflow
velocities using the Doppler effect formulae. $v_{\rm max}$ ($v_{\rm min}$) is the 
outflow velocity corresponding 
to the shortest (longest) wavelength associated with the 
\ion{C}{4}~$\lambda$1549 mini-BAL troughs. For quasars with a single absorption trough, 
 $v_{\rm max}$ and $v_{\rm min}$ mark the boundaries of the absorption trough where
the $C'$ parameter in Equation (2) equals unity. To keep consistency with previous
works on BAL and mini-BAL quasars (e.g., Trump et~al. 2006; G09; Gibson et~al. 2009b), 
the rest wavelength used for 
\ion{C}{4} $\lambda$1549 is 1550.77${\rm \AA}$, which is the red component of 
the \ion{C}{4} doublet line (Verner et~al. 1996). Relevant properties of the 
\ion{C}{4} mini-BALs are shown in Table~2.

\section{X-Ray Data Analysis}
Source detections were performed with the {\sc wavdetect} algorithm (Freeman et al. 2002) 
using a detection threshold of $10^{-4}$ and wavelet scales of $1$, $\sqrt{2}$, $2$, 
$2\sqrt{2}$ and $4$ pixels; all twelve {\it Chandra} targets were detected within 
$0.6''$ of the object's SDSS coordinates. Aperture photometry was performed 
for each object using IDL {\sc aper} routine with an aperture radius of $3''$ and 
inner and outer annulus radii of $6''$ and $9''$ for background subtraction, 
respectively. The resulting counts in the observed-frame soft ($0.5-2.0$ keV), 
hard ($2.0-8.0$ keV) and full ($0.5-8.0$ keV) \hbox{X-ray} bands are reported in Table~3.
Errors on the \hbox{X-ray} counts were calculated using Poisson 
statistics corresponding to the 1$\sigma$ significance level according to 
Tables 1 and 2 of Gehrels (1986). The band ratios and the effective power-law 
photon indices are also included in Table~3. The band ratio is defined as the number of 
hard-band counts divided by the number of soft-band counts. The errors on the band ratio 
correspond to the 1$\sigma$ significance level and were calculated using 
equation (1.31) in \S 1.7.3 of Lyons (1991). The band ratios for all of 
the {\it Chandra} objects can be directly compared with one another, because 
they were all observed during the same cycle. The effective power-law photon index
is the photon index of the presumed Galactic-absorbed power-law spectral model for 
each source, under which the band ratio would be the value calculated from the X-ray 
counts. It was calculated using the {\it Chandra} PIMMS\footnote{http://cxc.harvard.edu/toolkit/pimms.jsp} tool (version 3.9$k$), using the 
{\it Chandra} Cycle 10 instrument response to account for the 
decrease in instrument efficiency over time.
\begin{deluxetable*}{clccccc}
\tablecolumns{7} \tablenum{3} \tabletypesize{\footnotesize}
\tablewidth{\textwidth}
\tablecaption{X-Ray Counts}
\tablehead{
    \colhead{} & \colhead{}                     
    & \colhead{Full Band}     
    & \colhead{Soft Band}     
    & \colhead{Hard Band}     
    & \colhead{Band}   
    & \colhead{}\\
    \colhead{} & \colhead{Object Name (SDSS J)} 
    & \colhead{(0.5--8.0 keV)} 
    & \colhead{(0.5--2.0 keV)} 
    & \colhead{(2.0--8.0 keV)} 
    & \colhead{Ratio}  
    & \colhead{$\Gamma$}
}
\startdata
\multicolumn{2}{l}{{\it Chandra} Cycle 10 Objects} & & & & & \\
& $080117.79+521034.5$ & $23.9^{+6.0}_{-4.9}$ & $20.2^{+5.6}_{-4.4}$ 
& $3.8^{+3.1}_{-1.9}$  & $0.19^{+0.16}_{-0.10}$       & $2.3^{+0.7}_{-0.6}$ \\
& $091342.48+372603.3$ & $63.8^{+9.0}_{-8.0}$ & $38.8^{+7.3}_{-6.2}$ 
& $25.0^{+6.1}_{-5.0}$ & $0.64^{+0.20}_{-0.16}$       & $1.1^{+0.3}_{-0.2}$ \\
& $092914.49+282529.1$ & $33.0^{+6.8}_{-5.7}$ & $30.0^{+6.5}_{-5.4}$ 
& $3.0^{+2.9}_{-1.6}$ & $0.10^{+0.10}_{-0.06}$        & $2.8^{+0.7}_{-0.6}$ \\
& $093207.46+365745.5$ & $59.5^{+8.8}_{-7.7}$ & $53.8^{+8.4}_{-7.3}$ 
& $5.7^{+3.5}_{-2.3}$ & $0.11^{+0.07}_{-0.05}$        & $2.7^{+0.5}_{-0.4}$ \\
& $105158.74+401736.7$ & $209.8^{+15.5}_{-14.5}$ & $168.2^{+14.0}_{-13.0}$ 
& 41.6$^{+7.5}_{-6.4}$ & $0.25^{+0.05}_{-0.04}$ & $2.0^{+0.2}_{-0.2}$ \\
& $105904.68+121024.0$ & $44.6^{+7.7}_{-6.7}$ & $34.8^{+7.0}_{-5.9}$ 
& $9.8^{+4.2}_{-3.1}$ & $0.28^{+0.13}_{-0.10}$        & $1.9^{+0.4}_{-0.4}$ \\
& $120331.29+152254.7$ & $94.5^{+10.8}_{-9.7}$ & $76.6^{+9.8}_{-8.7}$ 
& $17.9^{+5.3}_{-4.2}$ & $0.23^{+0.08}_{-0.06}$      & $2.0^{+0.3}_{-0.3}$ \\
& $123011.84+401442.9$ & $81.7^{+10.1}_{-9.0}$& $60.8^{+8.8}_{-7.7}$ 
& $20.9^{+5.6}_{-4.5}$ & $0.34^{+0.11}_{-0.09}$       & $1.7^{+0.3}_{-0.2}$ \\
& $125230.84+142609.2$ & $53.0^{+8.3}_{-7.3}$ & $40.0^{+7.4}_{-6.3}$ 
& $13.0^{+4.7}_{-3.5}$ & $0.32^{+0.13}_{-0.10}$       & $1.7^{+0.3}_{-0.3}$ \\
& $141028.14+135950.2$ & $65.3^{+9.1}_{-8.1}$ & $50.8^{+8.2}_{-7.1}$ 
& $14.9^{+5.0}_{-3.8}$ & $0.29^{+0.11}_{-0.09}$       & $1.8^{+0.3}_{-0.3}$ \\
& $151451.77+311654.0$ & $22.6^{+5.8}_{-4.7}$ & $19.8^{+5.5}_{-4.4}$ 
& $2.8^{+2.9}_{-1.6}$ & $0.14^{+0.15}_{-0.08}$        & $2.5^{+0.8}_{-0.7}$ \\
& $224649.29-004954.3$ & $75.4^{+9.7}_{-8.7}$ & $58.0^{+8.7}_{-7.6}$ 
& $17.4^{+5.3}_{-4.1}$ & $0.30^{+0.10}_{-0.08}$       & $1.9^{+0.3}_{-0.3}$ \\ \\
\multicolumn{2}{l}{Archival {\it XMM-Newton} Objects} & & & & & \\
& $142656.17+602550.8$ & $237.0^{+16.4}_{-15.4}$ 
& $167.0^{+14.0}_{-12.9}$ & $53.0^{+8.3}_{-7.3}$ & $0.32^{+0.06}_{-0.05}$ 
& $1.8^{+0.1}_{-0.1}$ \\
& $160222.72+084538.4$\tablenotemark{a} & $16.0^{+5.1}_{-4.0}$ & $13.4^{+4.8}_{-3.6}$ 
& $2.6^{+2.8}_{-1.5}$  & \nodata      & \nodata
\enddata
\tablenotetext{a}{The counts for SDSS J$160222.72+084538.4$ are converted from the
counts in the 2XMM catalog, using an absorbed power-law
model with $\Gamma=2$ and the Galactic $N_H$ in Column~5 of Table~4.}
\end{deluxetable*}

The key \hbox{X-ray}, optical and radio properties of our sample are listed in Table~4. 

\noindent Column (1): The SDSS J2000 equatorial coordinates for the quasar.

\noindent Column (2): The quasar's redshift. These values were taken from the improved
measurements in Hewett \& Wild (2010), which have significantly reduced systematic 
biases compared to the redshift values in the SDSS DR5 quasar catalog. 

\noindent Column (3): The apparent $i$-band magnitude of the quasar using the 
SDSS quasar catalog BEST photometry, $m_{i}$.

\noindent Column (4): The absolute $i$-band magnitude for the quasar, $M_{i}$, from the 
SDSS DR5 quasar catalog, which was calculated assuming a power-law continuum index
of $\alpha_\nu\;=\;-0.5$.

\noindent Column (5): The Galactic neutral hydrogen column density obtained 
with the {\it Chandra} COLDEN tool, in units of $10^{20}$~cm$^{-2}$.

\noindent Column (6): The count rate in the observed-frame soft \hbox{X-ray} band 
($0.5-2.0$~keV), in units of $10^{-3}$~s$^{-1}$.

\noindent Column (7): The Galactic extinction-corrected flux in the 
observed-frame soft \hbox{X-ray} band ($0.5-2.0$~keV) obtained with
PIMMS, in units of $10^{-14}$~erg~cm$^{-2}$~s$^{-1}$. An 
absorbed power-law model was used with a photon index $\Gamma=2$, which is typical 
for quasars and is consistent with joint \hbox{X-ray} spectral fitting (see \S 4.1). 
The Galactic neutral hydrogen column density was used for each quasar 
($N_H$, given in Column~5 of Table~4).

\noindent Column (8): The flux density at rest-frame 2 keV obtained with 
PIMMS, in units of $10^{-32}$ erg cm$^{-2}$ s$^{-1}$ Hz$^{-1}$. Our sources
are generally brighter in 2~keV X-ray flux density than the mini-BAL quasars in G09 
(see Fig.~1$c$). 

\noindent Column (9): The logarithm of the Galactic extinction-corrected quasar 
luminosity in the rest-frame $2-10$~keV band obtained with PIMMS. 

\noindent Columns (10) and (11): The flux density at rest-frame 2500~\AA~ in 
units of $10^{-27}$ erg~cm$^{-2}$~s$^{-1}$~Hz$^{-1}$ and the logarithm of the 
monochromatic luminosity at rest-frame 2500~\AA, respectively. The flux density 
was obtained using a method similar to that described in \S 2.2 
of Vignali et al.~(2003). It is calculated as an interpolation between the fluxes 
of two SDSS filters bracketing rest-frame 2500~\AA. A bandpass correction is applied 
when converting flux into monochromatic luminosity.

\noindent Column (12): The \hbox{X-ray}-to-optical power-law slope, given by
\begin{equation}
    \alpha_{\rm ox} = \frac{{\rm log}(f_{\rm 2\;keV} / 
    f_{2500\mbox{\rm~\scriptsize\AA}})}{{\rm log}(\nu_{\rm 2\;keV} / \nu_{2500\mbox{\rm~\scriptsize\AA}})}.
\end{equation}
The value of $\alpha_{\rm ox}$ for SDSS J$142656.17+602550.8$ was taken 
directly from Shemmer et al.~(2008). 

\noindent Column (13): $\Delta\alpha_{\rm ox}$, defined as
\begin{equation}
    \Delta\alpha_{\rm ox} = \alpha_{\rm ox(measured)} - \alpha_{\rm ox(expected)}.
\end{equation}
The expected value of $\alpha_{\rm ox}$ for a typical quasar is calculated 
from the established $\alpha_{\rm ox}$-$L_{2500\mbox{\rm~\scriptsize\AA}}$ correlation given as 
Equation (3) of Just et al.~(2007). The statistical significance of this 
difference, given in parentheses, is in units of $\sigma$, where $\sigma$ is 
given in Table~5 of Steffen et al.~(2006) as the RMS $\alpha_{\rm ox}$ for 
various luminosity ranges. Here, $\sigma$~=~0.146 for 
31~$<$~log$L_{2500\mbox{\rm~\scriptsize\AA}}$~$<$~32 and $\sigma$~=~0.131 for 
32~$<$~log$L_{2500\mbox{\rm~\scriptsize\AA}}$~$<$~33.

\noindent Column (14): The radio-loudness parameter, given by 
\begin{equation}
    R = \frac{f_{\rm 5\;GHz}}{f_{4400\mbox{\rm~\scriptsize\AA}}}{\rm .}
\end{equation}
The denominator, $f_{4400\mbox{\rm~\scriptsize\AA}}$, was found via extrapolation from 
$f_{2500\mbox{\rm~\scriptsize\AA}}$ using an optical/UV power-law slope of $\alpha_\nu$ = $-0.5$. 
The numerator, $f_{5\;{\rm GHz}}$, was found using a radio power-law slope of 
$\alpha_\nu$ = $-0.8$ and a flux at 20 cm, $f_{20\;{\rm cm}}$, of three times the 
RMS noise in a $0.5'\times0.5'$ {\it FIRST} image cutout at the object's 
coordinates. All the quantities of flux density here
are per unit frequency. This value of $f_{20\;{\rm cm}}$ was used as an upper limit 
because there were no {\it FIRST} detections. $R$ is thus 
listed as an upper limit for all of the objects. All the upper limits are well 
below the commonly used criterion for radio-quiet objects ($R < 10$). Therefore 
any \hbox{X-ray} emission from jets can be neglected. 


\section{Discussion}

\subsection{Joint \hbox{X-ray} Spectral Analysis}
\begin{deluxetable*}{ccccccccc}
\tablecolumns{9} \tablenum{5} \tabletypesize{\scriptsize}
\tablewidth{\textwidth}
\tablecaption{X-ray Spectral Analysis}
\tablehead{
\colhead{}
    & \colhead{}
    & \colhead{}
    & \multicolumn{2}{c}{Power Law}
    & \colhead{}
    & \multicolumn{3}{c}{Power Law}\\
    \colhead{}
    & \colhead{Total Full-Band}
    & \colhead{}
    & \multicolumn{2}{c}{with Galactic Absorption}
    & \colhead{}
    & \multicolumn{3}{c}{with Galactic and Intrinsic Absorption}
    \\
    \cline{4-5}\cline{7-9}\\    
\colhead{Sources} 
    & \colhead{Counts} 
    & \colhead{}
    & \colhead{$\Gamma$} 
    & \colhead{$C$-Statistic (bins\tablenotemark{a})} 
    & \colhead{}
    & \colhead{$\Gamma$} 
    & \colhead{$N_H (10^{22}{\rm cm}^{-2})$}
    & \colhead{$C$-Statistic (bins\tablenotemark{a})}
}
\startdata
SDSS J105158.74+401736.7 & 210 & & $1.99^{+0.21}_{-0.21}$ & $106.17\ (111)$ & & $2.06^{+0.37}_{-0.26}$ & $<1.77$ & $106.02\ (111)$ \\
All 12 {\it Chandra} Snapshot Survey Objects & 815 & & $1.90^{+0.11}_{-0.11}$ & $523.18\ (599)$ & & $1.91^{+0.14}_{-0.11}$ & $<0.82$ & $523.17\ (599)$
\enddata
\tablenotetext{a}{The number of bins is smaller than the number of total counts because of the grouping of the spectra (see \S 4.1).}
\label{tab5}
\end{deluxetable*}
Constraining \hbox{X-ray} spectral properties, especially the power-law photon index and 
the intrinsic absorption column density, can shed light on the \hbox{X-ray} 
generation mechanisms and the nuclear environments of mini-BAL quasars. We investigate 
the \hbox{X-ray} spectra of the 12 mini-BAL quasars observed in the {\it Chandra} snapshot
survey. Joint spectral fitting is performed for the 12 sources, most of which 
do not have sufficient counts for individual analysis. We only did spectral fitting 
individually for one source (SDSS~J1051+4017) which has $\approx 210$ counts. Excluding
this source in the joint fitting does not significantly change the best-fit 
parameters, which indicates this source does not introduce biases into the 
joint fitting. The \hbox{X-ray} spectra were extracted with the CIAO routine {\sc psextract} 
using an aperture of $3''$ radius centered on the \hbox{X-ray} position for each source. 
Background spectra were extracted using an annulus with an inner radius of $6''$ 
and outer radius of $9''$. All background regions are free of \hbox{X-ray} sources. 

Spectral analysis was carried out with XSPEC v12.5.1 (Arnaud 1996) using the $C$-statistic 
(Cash 1979). The X-ray spectrum for each source was grouped to have at least one count 
per energy bin. This grouping procedure
avoided zero-count bins while allowing retention of all 
spectral information. The $C$-statistic is well suited to the low-count scenario of our 
analysis (e.g., Nousek \& Shue 1989). Each source is assigned its own redshift and 
Galactic column density in the joint fitting. The following models were 
employed in the spectral fitting: (1) a power-law model with a Galactic absorption 
component represented by the \verb+wabs+ model in XSPEC (Morrison \& McCammon 1983), 
in which the Galactic column density is fixed to the values from 
COLDEN (see Column~5 in Table~4); (2) a model similar to the first, but 
adding an intrinsic (redshifted) neutral absorption component, represented by the \verb+zwabs+ 
model. The errors or the upper limits for the best-fit spectral parameters are quoted at 90\% 
confidence level for one parameter of interest ($\Delta C=2.71$; Avni 1976; 
Cash 1979).

Table~5 shows the \hbox{X-ray} spectral fitting results. For the individually fitted 
source, SDSS J1051+4017, the best-fit photon index is $\Gamma=1.99^{+0.21}_{-0.21}$
without adding intrinsic absorption. This value is consistent with that from
band-ratio analysis (see Table~3). After adding the intrinsic absorption component, 
the photon index changes insignificantly to $\Gamma=2.06^{+0.37}_{-0.26}$. The \hbox{power-law} becomes
slightly steeper. The constraint on the intrinsic column density is weak
($N_H \lesssim 1.8\times 10^{22} \ {\rm cm}^{-2}$). The spectrum of this source is shown 
in Fig.~3{\it (a)}, binned to have at least 10 counts per bin for the purpose of presentation. 
For the joint fitting, the mean photon index is $\Gamma=1.90^{+0.11}_{-0.11}$ in the
model without intrinsic absorption. After adding the intrinsic absorption component, 
the mean photon index changes slightly to  $\Gamma=1.91^{+0.14}_{-0.11}$. This result
is consistent with the absence of evidence for strong intrinsic neutral absorption 
on average (\hbox{$N_H \lesssim 8.2\times 10^{21} \ {\rm cm}^{-2}$}). 
An \hbox{$F$-test} shows that adding an intrinsic absorption component does not significantly 
improve the fit quality ($\Delta C=0.15$ for SDSS~J1051+4017; \hbox{$\Delta C=0.01$} for the
joint fitting). The \hbox{{\it XMM-Newton}} archival source, SDSS J1426+6025, has 
similar \hbox{X-ray} spectral properties (\hbox{$\Gamma=1.76^{+0.14}_{-0.13}$}, 
\hbox{$N_H \lesssim 2.3\times 10^{22} \ {\rm cm}^{-2}$;} Shemmer et al. 2008). 
The lack of substantial intrinsic absorption in these \hbox{mini-BAL} quasar 
spectra shows that mini-BAL quasars are more similar to non-BMB quasars in their
\hbox{X-ray} absorption properties. Fig.~3{\it (b)} shows a contour plot of the $\Gamma-N_H$ parameter 
space for the joint fitting with intrinsic absorption added. 

We also investigated whether an ionized intrinsic absorber could better 
describe the \hbox{X-ray} spectra of the mini-BAL quasars. The \verb+absori+ model 
(Done et al. 1992), instead of the previously used \verb+zwabs+ model, was used 
to represent intrinsic absorption. The \verb+absori+ model has more spectral parameters, 
and thus requires higher-quality X-ray spectra for the placement of useful spectral constraints. 
However, due to the limited counts of the 
sources in our sample, the \hbox{X-ray} spectral fitting could not provide a preference 
between the neutral and ionized intrinsic absorption models. 



\subsection{Relative \hbox{X-ray} Brightness}

The $\Delta\alpha_{\rm ox}$ parameter, defined by Equation (4), quantifies the relative
\hbox{X-ray} brightness of a mini-BAL quasar with respect to a typical non-BMB quasar
with the same UV luminosity. Fig.~4 and Fig.~5 show the distributions of 
$\alpha_{\rm ox}$ and $\Delta\alpha_{\rm ox}$, respectively, for our sample, as 
well as the \hbox{non-BMB}, mini-BAL and HiBAL quasars in G09. The mean values of 
$\alpha_{\rm ox}$ and $\Delta\alpha_{\rm ox}$ were calculated using the 
Kaplan-Meier estimator implemented in the Astronomy Survival Analysis (ASURV) 
package (e.g., Lavalley et al. 1992). A filled 
triangle with horizontal error bars is used to show the mean value of 
$\alpha_{\rm ox}$ and $\Delta\alpha_{\rm ox}$ for our sample combined with the 
G09 mini-BAL quasars. The mean values for $\alpha_{\rm ox}$ 
and $\Delta\alpha_{\rm ox}$ for the non-BMB and HiBAL quasars in the G09 
sample only are shown using a filled square with horizontal error 
bars. 

After our sample is combined with the mini-BAL quasars in G09, 
the mean $\alpha_{\rm ox}$ value, $-$1.68~$\pm$~0.02, 
remains the same (with a smaller error bar) as the value for the mini-BAL quasar 
sample in G09 only. The mean $\Delta\alpha_{\rm ox}$ value became slightly less 
negative once our sample was included: the mean $\Delta\alpha_{\rm ox}$ value of 
our sample combined with the mini-BAL quasar sample from G09 is 
$-$0.03~$\pm$~0.02, while it was $-$0.05~$\pm$~0.03 for the G09 sample.
Therefore, the mean $\Delta\alpha_{\rm ox}$ of the combined mini-BAL quasar sample is
closer to that of the non-BMB quasars in G09 ($0.00$~$\pm$~0.01), which 
strengthens the trend that mini-BAL quasars are only slightly \hbox{X-ray} weaker than non-BMB 
quasars, while BAL quasars are considerably \hbox{X-ray} weaker than both mini-BAL and non-BMB quasars 
(the mean $\Delta\alpha_{\rm ox}$ of the BAL quasar sample in G09 is 
$-$0.22~$\pm$~0.04). 

We have performed two-sample tests to assess if the $\Delta\alpha_{\rm ox}$ values for mini-BAL 
quasars follow the
same distribution as those for BAL quasars or non-BMB quasars. Gehan tests (Gehan 1965), also 
implemented in the ASURV package, were used 
on the distributions of $\Delta\alpha_{\rm ox}$ values for the combined mini-BAL quasar sample, 
and the non-BMB and BAL quasar samples in G09. The $\Delta\alpha_{\rm ox}$ distributions
of mini-BAL and non-BMB quasars are not found to be different at a high level of significance; 
the probability of the null hypothesis from the Gehan test is 19.3\%. In contrast, 
the $\Delta\alpha_{\rm ox}$ distributions of mini-BAL and BAL quasars clearly differ, 
with a null-hypothesis probability of 0.01\%. Sources with upper limits on 
$\Delta\alpha_{\rm ox}$ are included in the two sample tests. 

We have tested the normality of the distribution of $\Delta\alpha_{\rm ox}$ for mini-BAL quasars. 
Deviations from a normal distribution, such as a skew tail, 
may indicate a sub-population of physically special sources (e.g., extremely \hbox{X-ray} 
weak sources) in our sample. Anderson-Darling tests (e.g., Stephens 1974) were performed 
on the $\Delta\alpha_{\rm ox}$ values of the mini-BAL quasars. 
First we examine the distribution of the 14 sources in our sample. 
The Anderson-Darling test cannot reject the hypothesis that the $\Delta\alpha_{\rm ox}$ values 
follow a Gaussian distribution; the rejection probability is only 0.5\%. 
Adding the mini-BAL quasars in G09, the sample size increases to 61. However, 
there are 8 sources having upper limits on their $\Delta\alpha_{\rm ox}$ values. Since the
Anderson-Darling test (as well as other commonly used normality tests) can only treat 
uncensored data, we generate a sample of 40 sources in an unbiased way with all sources 
detected by considering only the mini-BAL quasars 
observed by {\it Chandra} with an exposure longer than 2~ks. One mini-BAL quasar in G09, 
SDSS~J113345.62+005813.4, which is undetected in a {\it Chandra} exposure of 4.3~ks, is removed from this
sample because this source likely has both intervening and intrinsic X-ray absorption 
(Hall et al. 2006) while we are investigating intrinsic X-ray absorption only. 
The probability of non-normality of the $\Delta\alpha_{\rm ox}$ values 
for this sample is only 3.6\%. Therefore, the $\Delta\alpha_{\rm ox}$ values of mini-BAL quasars
follow the Gaussian distribution to a reasonable level of approximation. 
Strateva et al.~(2005) reported that for their \hbox{non-BMB} quasar
sample (with only a few percent BAL quasar contamination), a Gaussian profile could 
provide a reasonable fit to the distribution of the $\Delta\alpha_{\rm ox}$ values. Gibson
et al.~(2008b) also found that normality of the $\Delta\alpha_{\rm ox}$ distribution for their
radio-quiet non-BMB quasar sample could not
be ruled out by the Anderson-Darling test. For our sample, we find no evidence that any 
significant subset of sources falls outside the Gaussian regime. Thus mini-BAL quasars are similar to non-BMB 
quasars with respect to the distribution of relative \hbox{X-ray} brightness. We also fit the
$\Delta\alpha_{\rm ox}$ distributions using the IDL {\sc gaussfit} procedure (see Fig.~6). 
For the 14 sources in our sample, 
the best-fit parameters are $\mu=0.04\pm0.01,\ \sigma=0.07\pm0.01$. For the 40 mini-BAL 
quasars observed by {\it Chandra} with an exposure longer than 2~ks, the fitting results are 
$\mu=0.03\pm0.02,\ \sigma=0.12\pm0.02$. The measurement errors for the $\Delta\alpha_{\rm ox}$ values
for most mini-BAL quasars are $\approx 0.03$, which is much smaller than the intrinsic spread of 
$\Delta\alpha_{\rm ox}$. In fact, the square roots of the $\Delta\alpha_{\rm ox}$ variances for the 
two groups of mini-BAL quasars tested above are 0.086 and 0.137, respectively, which indicates 
that the measurement errors have smaller influence than the spread of $\Delta\alpha_{\rm ox}$.  

\subsection{Mini-BAL Quasar Reddening}
\begin{deluxetable*}{rcrrcrrcrrcrr}
\tablecolumns{13} \tablenum{6} \tabletypesize{\footnotesize}
\tablewidth{\textwidth}
\tablecaption{Spearman Rank-Order Correlation Coefficients and Probabilities}
\tablehead{
     \colhead{} &  &\multicolumn{2}{c}{G09 Sample\tablenotemark{a}} & &\multicolumn{2}{c}{G09 Sample} & & \multicolumn{2}{c}{G09 Sample} & & \multicolumn{2}{c}{G09 mini-BAL only}\\
     \colhead{} &  &\multicolumn{2}{c}{}           & &\multicolumn{2}{c}{+ Our Sample} & & \multicolumn{2}{c}{+ Our Sample} & & \multicolumn{2}{c}{+ Our Sample}\\
     \colhead{} &  &\multicolumn{2}{c}{(1,000~km~s$^{-1}$)\tablenotemark{b}} & &\multicolumn{2}{c}{(1,000~km~s$^{-1}$)\tablenotemark{b}} & & \multicolumn{2}{c}{(500~km~s$^{-1}$)\tablenotemark{b}} & & \multicolumn{2}{c}{(500~km~s$^{-1}$)\tablenotemark{b}}
     \\
     \cline{3-4}\cline{6-7}\cline{9-10}\cline{12-13}\\
\colhead{} &  & \colhead{$\rho$} & \colhead{$1-P_S$} & &  \colhead{$\rho$} & \colhead{$1-P_S$} & & \colhead{$\rho$} & \colhead{$1-P_S$} & & \colhead{$\rho$} & \colhead{$1-P_S$}
}
\startdata
                      $AI$ vs. $\Delta\ \alpha_{\rm ox}$ & & $-0.474$ & $99.76\%$ & & $-0.519$ & $99.99\%$ & & $-0.599$ & $>99.99\%$ & & $-0.434$ & $98.59\%$ \\	
              $v_{\rm max}$ vs. $\Delta\ \alpha_{\rm ox}$ & & $-0.442$ & $99.54\%$ & & $-0.380$ & $99.52\%$ & & $-0.409$ & $99.76\%$  & & $-0.402$ & $97.69\%$ \\
              $v_{\rm min}$ vs. $\Delta\ \alpha_{\rm ox}$ & & $-0.014$ & $7.00\%$ & & $-0.055$ & $32.74\%$  & & $-0.189$ & $83.92\%$ & & $-0.252$ & $84.56\%$\\
              $\Delta v$ vs. $\Delta\ \alpha_{\rm ox}$ & & $-0.620$ & $99.99\%$ & & $-0.628$ & $>99.99\%$ & & $-0.463$ & $99.94\%$ & & $-0.422$ & $98.30\%$\\
              $v_{\rm wt}$ vs. $\Delta\ \alpha_{\rm ox}$ & & $-0.191$ & $78.99\%$ & & $-0.205$ & $87.24\%$ & & $-0.252$ & $93.85\%$ & & $-0.260$ & $85.87\%$
\enddata
\tablenotetext{a}{The correlation analysis for G09 sample sources only is repeated with the Spearman rank-order correlation analysis, using updated $AI$, $v_{\rm max}$, $v_{\rm min}$, and $\Delta v$ values according to the improved redshift measurements in Hewett \& Wild (2010).}
\tablenotetext{b}{Velocity width limit used when defining mini-BALs.}
\end{deluxetable*}\label{table6}
BAL quasars are found to be redder than non-BMB quasars (e.g., Weymann et al. 1991; Brotherton et al. 
2001; Reichard et al. 2003; Trump et al. 2006; Gibson et al. 2009b), which is likely caused by
increased dust reddening. Following Gibson et al. (2009b), we investigate the reddening of mini-BAL quasars
by measuring the quantity $R_{\rm red}$, defined as, 
\begin{equation}
R_{\rm red} \equiv F_{\nu}(1400 \ {\rm \AA})/F_{\nu}(2500 \ {\rm \AA}),
\end{equation}
where $F_{\nu}(1400 \ {\rm \AA})$ and $F_{\nu}(2500 \ {\rm \AA})$ are the continuum flux densities 
at rest-frame $1400$ \AA~ and $2500$~\AA, respectively. These two wavelengths approximately represent the broadest 
available spectral coverage for most of our mini-BAL quasars. The continuum flux densities were calculated 
by fitting the SDSS spectra using the algorithm described in \S 2.1 of Gibson et al.~(2008a).

Fig.~7 shows the distribution of $R_{\rm red}$ values for the 61 mini-BAL quasars from this work and G09. 
The median value of $R_{\rm red}$ is 0.70. Gibson et al. (2009b) studied the $R_{\rm red}$ values of 
\hbox{non-BMB} and BAL quasars in the redshift range $1.96\leq z\leq2.28$, and found median $R_{\rm red}$ values 
for non-BMB, HiBAL and LoBAL quasars of 0.75, 0.64,
and 0.30, respectively. The median $R_{\rm red}$ of mini-BAL quasars is intermediate between 
those of non-BMB and HiBAL quasars. We performed
Kolmogorov-Smirnov tests on the distribution of $R_{\rm red}$ values for mini-BAL quasars versus
those for non-BMB, HiBAL and LoBAL quasars in the common redshift range of $1.96\leq z\leq2.28$. 
The probability of the $R_{\rm red}$ values for 27 mini-BAL quasars and for 6,689 non-BMB quasars 
coming from the same population is 0.100, while those probabilities are 0.053 and $3.05\times10^{-8}$ 
for the mini-BAL quasars compared to 904 HiBAL quasars and 82 LoBAL quasars, respectively. These
results cannot distinguish whether the $R_{\rm red}$ distribution for mini-BAL quasars is more similar to that
for non-BMB quasars or for HiBAL quasars.

The ``redness'' of the mini-BAL quasars may be caused by either steeper intrinsic 
\hbox{power-law} slopes or dust reddening. 
We calculated the relative SDSS colors $\Delta(u-r)$, $\Delta(g-i)$, and $\Delta(r-z)$, to assess curvature in 
the optical continuum, which can be seen as the evidence of dust reddening (e.g., Hall et al. 2006). 
The relative color is defined as the difference between the source color and the modal color for SDSS quasars 
in the corresponding redshift bin, calculated following the method in \S 5.2 of Schneider et~al.~(2007). 
For a pure power-law continuum, 
these three relative colors should be equal within uncertainties, while for dust-reddened objects, 
we would expect curvature in the continuum, specifically, $\Delta(u-r)\;>\;\Delta(g-i)\;>\;\Delta(r-z)$. 
For our mini-BAL quasar sample, the median values 
\footnote{The quoted error for the median value is estimated using $NMAD/\sqrt{N}$. $NMAD$, 
the normalized median absolute deviation, is a robust estimator of the spread of the sample, defined as 
\hbox{$NMAD=1.48\times{\rm median}(|x-{\rm median}(x)|)$} (see \S 1.2 of Maronna~et~al.~2006). $N$ is the sample
size, which is 61 in our case.}
of the three relative colors are \hbox{$0.086\pm0.025$},
\hbox{$0.074\pm0.018$}, and \hbox{$0.032\pm0.015$}, respectively. Kolmogorov-Smirnov tests show that the probability
for $\Delta(u-r)$ and $\Delta(r-z)$ following the same distribution is $7.5\times10^{-3}$, while those probabilities for 
 $\Delta(u-r)$ vs. $\Delta(g-i)$, and  $\Delta(g-i)$ vs. $\Delta(r-z)$ are 0.35 and 0.068, respectively. Therefore, 
we can see a moderate trend that 
\hbox{$\Delta(u-r)\;>\;\Delta(g-i)\;>\;\Delta(r-z)$}, indicating mild dust reddening in \hbox{mini-BAL} quasar spectra.

Assuming all quasars
have a similar underlying continuum shape on average, different $R_{\rm red}$ values represent different
dust-reddening levels. According to an SMC-like dust extinction curve (Pei 1992), the median  
$R_{\rm red}$ for mini-BAL quasars corresponds to $E(B-V)\approx0.011$, which is lower than the $E(B-V)$ values of HiBAL
quasars ($\approx0.023$) and LoBAL quasars ($\approx0.14$) in G09. Since the relative \hbox{X-ray} brightness
$\Delta\alpha_{\rm ox}$ largely represents the absorption in \hbox{X-ray} band, we tested for correlations 
between $R_{\rm red}$ and $\Delta\alpha_{\rm ox}$ for the
mini-BAL quasars using the Spearman rank-order correlation analysis.
This analysis returns a correlation coefficient $\rho$ and a corresponding probability $P_S$ for the null
hypothesis (no correlation). $\rho=0$ indicates no correlation between the two variables tested, while 
$\rho=\pm1$ means a perfect correlation. The correlation probability is $1-P_S$, e.g.,
$P_S=0.01$ means a correlation probability of 99\%. We find a marginal correlation between 
$R_{\rm red}$ and $\Delta\alpha_{\rm ox}$ for 
the 61 mini-BAL quasars in the combined sample at an 86.2\% probability \hbox{($\rho=0.190$)}.

\subsection{UV Absorption vs. \hbox{X-ray} Brightness}
\begin{deluxetable*}{clrrrrcl}
\tablecolumns{8} \tablenum{7} \tabletypesize{\scriptsize}
\tablewidth{\textwidth}
\tablecaption{\ion{C}{4} Mini-BAL Properties for Sources with New Absorption Troughs
Included under the Definition of $AI_{500}$}
\tablehead{
    \colhead{} & \colhead{}                     & \colhead{$AI_{500}$}         
    & \colhead{$v_{\rm min}$} & \colhead{$v_{\rm max}$} & \colhead{$v_{\rm wt}$} 
    & \colhead{No. of Troughs} & \colhead{}\\ 
    \colhead{} & \colhead{Object Name (SDSS J)} & \colhead{(km~s$^{-1}$)} 
    & \colhead{(km~s$^{-1}$)} & \colhead{(km~s$^{-1}$)}  & \colhead{(km~s$^{-1}$)} 
    & \colhead{Contributing to $AI$} & \colhead{Notes\tablenotemark{a}}
}
\startdata
\multicolumn{2}{l}{{\it Chandra} Cycle 10 Objects} & & & \\
& $091342.48+372603.3$ & $1136.8$  & $3179$ & $23014$ & $10715$ & 4 & Feature at 
                                                                $\approx$1535~
                                                                \AA~ included\\
& & & & & & & Feature at $\approx$1480 ~\AA~ included\\
& $105904.68+121024.0$ & $2503.5$  & $5378$ & $23349$ & $11717$ & 6 & Feature at 
                                                                $\approx$1470~
                                                                \AA~ included\\
& $151451.77+311654.0$ & $692.7$  & $24124$ & $27003$ & $25592$ & 2 & Feature at 
                                                                $\approx$1430~
                                                                \AA~ included\\ \\
\multicolumn{2}{l}{Archival {\it XMM-Newton} Objects} & & & \\
& $142656.17+602550.8$ & $478.3$  & $4937$ & $28324$   & $21746$ & 2 & Feature at 
                                                                $\approx$1525~
                                                                \AA~ included\\
& $160222.72+084538.4$ & $517.1$  & $2136$ & $15239$   & $10006$ & 2 &  Feature at 
                                                                $\approx$1540~
                                                                \AA~ included
\enddata
\tablenotetext{a}{This column describes the newly included absorption troughs using the 
velocity width limit of $500$~km~s$^{-1}$. As discussed in the ``Notes" column of Table~2, 
some of them failed the width criterion when the velocity width limit was taken as 
 $1,000$~km~s$^{-1}$.}
\end{deluxetable*}
We define a high signal-to-noise ratio ($S/N$) sample of BAL and mini-BAL quasars to 
study correlations between $\Delta\alpha_{\rm ox}$ and  $AI$, $v_{\rm max}$, $v_{\rm min}$, 
$\Delta v$ ($\equiv|v_{\rm max}-v_{\rm min}|$) (see \S 2 for the definitions of these quantities). 
Radio-quiet BAL and mini-BAL quasars in G09 with 
$SN_{1700} > 9$, and the 14 mini-BAL quasars analyzed in this paper (which also have $SN_{1700} > 9$), 
are included in the high $S/N$
sample. $SN_{1700}$ is defined as the median of the flux divided by the noise
provided by the SDSS pipeline for spectral bins between rest-frame 1650~\AA~ and 1700~\AA. 
Excluding low $S/N$ sources enables the most-reliable measurements for the UV absorption 
parameters listed above; Gibson~et~al.~(2009b) reported that high $S/N$ spectra enabled more accurate 
identifications of broad absorption features. The values of the $AI$, $v_{\rm max}$, and $v_{\rm min}$
parameters for the sources in G09 are updated according to the improved redshift measurements
in Hewett \& Wild (2010). There are 23 BAL quasars (4 of them have upper limits on 
$\Delta\alpha_{\rm ox}$) and 33 mini-BAL quasars (2 of them have upper limits on 
$\Delta\alpha_{\rm ox}$) in the high $S/N$ sample. 

Fig.~8 shows the SDSS spectra for the 33 mini-BAL quasars 
in the high $S/N$ sample in order of $\Delta\alpha_{\rm ox}$. The positions of mini-BAL troughs
are shown in the figure. Some mini-BAL troughs show doublet-like features.
However, for mini-BAL quasars only, no significant trends are apparent between 
$\Delta\alpha_{\rm ox}$ and the positions or morphologies of mini-BAL troughs. 

In Fig.~9 we plot the UV absorption properties against $\Delta\alpha_{\rm ox}$ for 
the high $S/N$ mini-BAL and BAL quasar sample. The correlation probabilities between $\Delta\alpha_{\rm ox}$ and $AI$, 
$v_{\rm max}$, $v_{\rm min}$, $\Delta v$ are determined using 
Spearman rank-order correlation analysis (see Table~6). Our correlations are consistent with those found in G09. 
Our enhanced correlation between $AI$ and $\Delta\alpha_{\rm ox}$ confirms the 
trend found in G09 that UV broad absorption strength decreases with the relative \hbox{X-ray}
brightness. As in G09, $v_{\rm max}$ is correlated with 
$\Delta\alpha_{\rm ox}$, while there is no significant correlation
between $v_{\rm min}$ and $\Delta\alpha_{\rm ox}$. The $\Delta v$-$\Delta\alpha_{\rm ox}$ 
correlation shows that mini-BAL quasars with narrower absorption troughs are 
relatively \hbox{X-ray} brighter than BAL quasars. Some mini-BAL quasars have large $\Delta v$ values
because they have multiple absorption troughs. 

We have calculated the weighted average velocity, $v_{\rm wt}$, 
which is defined in Trump~et~al.~(2006) using the following equation,
\begin{equation}
v_{\rm wt}=\frac{\int_{0}^{29,000}(1-f(v))\ v\ C'dv}{AI},
\end{equation}
where $f(v)$ and $C'$ have the same definition as in Equation (2). The values of $v_{\rm wt}$ for the 14 \hbox{mini-BAL} 
quasars in our sample are listed in Table~2, and also shown in Fig.~8. The $v_{\rm wt}$ parameter represents 
the weighted average position of mini-BAL and BAL troughs. However, for some \hbox{mini-BAL} quasars with multiple absorption
troughs, $v_{\rm wt}$ may fall between troughs where no absorption features are present. 
$v_{\rm wt}$ is marginally correlated with $\Delta\alpha_{\rm ox}$ at a 
probability of 87.9\% for the high $S/N$ sample of mini-BAL and BAL quasars (see Fig.~9$e$).

In the disk-wind model for BAL and mini-BAL quasars (e.g., Murray et al. 1995), the UV absorbing outflow 
is radiatively accelerated, while
the material interior to the UV absorber shields the soft \hbox{X-ray} radiation to prevent overionization. 
This model may be supported by the correlations between UV absorption properties and relative \hbox{X-ray} weakness 
(see Table~6). As in G09, we also find some high 
velocity, relatively \hbox{X-ray} bright mini-BAL quasars in our newly analyzed 14 sources. These
sources can perhaps be fit into the ``failed'' BAL quasar scenario discussed in \S4.2 of G09, 
in which the high incident \hbox{X-ray} flux overionizes the outflow material so that BAL troughs cannot
form along the line of sight. 

\subsection{Alternative Measures of UV Absorption}

\subsubsection{Effects of the Velocity Width Limit}

From the rest-frame UV spectra of mini-BAL quasars (e.g., Fig.~2), we can see that some likely 
intrinsic absorption
features are missed by the definition of mini-BALs using the parameter $AI$ (also see the
``Notes'' column of Table~2). When calculating 
$AI$, only absorption troughs with velocity widths greater than $1,000$~km~s$^{-1}$ are included. 
This somewhat conservative velocity width limit was adopted because we wanted to spend valuable 
{\it Chandra} observing time only on sources with bona fide intrinsic absorption features.  
Here we now slightly modify the definition of $AI$ by lowering the velocity width limit to 
$500$~km~s$^{-1}$ to form a new parameter $AI_{500}$. This velocity width limit is similar to the 
value in the original introduction of $AI$ ($450$~km~s$^{-1}$) in Hall et al. (2002). The limit of 
$500$~km~s$^{-1}$ can include more broad absorption features while still minimizing the inclusion 
of, e.g., complex intervening systems (see the discussion in Hall et al. 2002). The equation for 
calculating $AI_{500}$ is the same as that of $AI$, 
\begin{equation}
AI_{500}=\int_{0}^{29,000}(1-f(v))\ C^*dv,
\end{equation}
while $C^*$ here has a different definition from $C'$ in Equation (2):
$C^*$~=~1 when the velocity width is at least 500~km~s$^{-1}$ and the absorption trough falls at 
least 10\% below the continuum; $C^*$~=~0 otherwise. Five of the 14 mini-BAL quasars in our sample 
have new absorption troughs included using this new velocity width limit (see the blue spectra in Fig.~8). 
The $AI_{500}$, as well as $v_{\rm max}$, $v_{\rm min}$, $\Delta v$, and $v_{\rm wt}$ also change accordingly (see Table~7). 

New absorption troughs are also identified for some BAL and mini-BAL quasars in G09 using the $AI_{500}$ definition 
(e.g., see Fig.~8 for mini-BAL quasars). From Fig.~8 we can see that more relatively
\hbox{X-ray} weak sources (with more negative $\Delta\alpha_{\rm ox}$) have new absorption troughs (indicated by
the red horizontal bars) included 
compared to relatively \hbox{X-ray} bright sources. We repeated the correlation analysis between 
$\Delta\alpha_{\rm ox}$ and UV absorption properties for the high $S/N$ BAL and mini-BAL quasar sample; 
the results are listed in Table~6. The UV
absorption properties under the new definition of $AI$ are plotted against $\Delta\alpha_{\rm ox}$ 
in Fig.~10. Red data points show the quantities changed under the new definition.   
All UV absorption parameters, except
$\Delta v$, have better correlations with $\Delta\alpha_{\rm ox}$, while the correlation 
probability between $\Delta\alpha_{\rm ox}$ and $\Delta v$ is still consistent with the previous result. 
$v_{\rm min}$ now has a marginal correlation (83.9\%) with 
$\Delta\alpha_{\rm ox}$. If we only consider mini-BAL quasars in the high $S/N$ sample (i.e., 
leaving out BAL quasars), these correlations 
still exist, although they are somewhat weaker (see the last two columns of Table~6). However, under the velocity
width limit of $1,000$~km~s$^{-1}$, the correlation between $\Delta\alpha_{\rm ox}$ and $AI$ disappears ($1-P_S=60.0\%$)
for mini-BAL quasars only, while the correlation between $\Delta\alpha_{\rm ox}$ and $v_{\rm max}$ becomes much 
weaker ($1-P_S=91.4\%$). 
We consider $500$~km~s$^{-1}$ to be a better velocity width limit than the 
previously used value ($1,000$~km~s$^{-1}$), in the sense that it can reveal somewhat clearer correlations
between \hbox{X-ray} and UV absorption properties. Fig.~8 shows that most of the newly included
absorption troughs have similar profiles and velocities to those troughs with $>1,000$~km~s$^{-1}$ velocity widths.
A Kolmogorov-Smirnov test does not distinguish the distributions of central velocities for the troughs with 
$>1,000$~km~s$^{-1}$ velocity widths and those with $500-1,000$~km~s$^{-1}$ velocity widths ($P=0.46$).
Therefore these newly included troughs are not likely to be caused by intervening systems.  

\subsubsection{Investigating New UV Absorption Parameters}
The $\Delta v$ parameter measures the velocity span of the UV absorbing outflow for sources 
with a single mini-BAL or BAL trough. However, for sources
with multiple absorption troughs, $\Delta v$ also includes regions where no absorption features are 
present. We therefore define a new parameter, called ``total velocity span ($v_{\rm tot}$)'', for the mini-BAL
and BAL quasars. $v_{\rm tot}$ is defined by the following equation, 
\begin{equation}
v_{\rm tot}=\int_{0}^{29,000} C^*dv,
\end{equation}
where $C^*$ is the same as that in the definition of $AI_{500}$. The values of $v_{\rm tot}$ for the 14 
mini-BAL quasars in our sample are listed in Table~8. The $v_{\rm tot}$ parameter represents 
the total velocity width of all absorption troughs up to $29,000$~km~s$^{-1}$. 
$v_{\rm tot}$ is correlated with $\Delta\alpha_{\rm ox}$ at a 
probability of $>99.99\%$ for the high $S/N$ sample of BAL and mini-BAL quasars (see Fig.~11$a$). This strong 
correlation also stands for mini-BAL quasars only ($1-P_s=99.77\%$).

As mentioned in \S 4.4, the weighted average velocity, $v_{\rm wt}$, may represent regions without 
absorption features for spectra with multiple troughs. Therefore we also
calculated the weighted average velocity for the strongest absorption trough $v_{\rm wt,st}$ using the 
following equation similar to the definition of $v_{\rm wt}$, 
\begin{equation}
v_{\rm wt,st}=\frac{\int_{v_{\rm min,st}}^{v_{\rm max, st}}(1-f(v))\ v\ dv}{AI_{\rm st}},
\end{equation}
where $v_{\rm min, st}$ and $v_{\rm max, st}$ are the minimum and maximum velocity for the strongest absorption
trough, which has the largest $AI$ value (denoted by $AI_{\rm st}$) among those multiple troughs. The $v_{\rm wt,st}$ 
values for the 14 mini-BAL quasars in our sample are listed in Table~8. $v_{\rm wt,st}$ has a better correlation 
(98.9\%) with $\Delta\alpha_{\rm ox}$ than $v_{\rm wt}$ (see Fig.~11$b$).
\begin{deluxetable}{clrrr}
\tablecolumns{5} \tablenum{8} \tabletypesize{\footnotesize}
\tablewidth{0pt}
\tablecaption{The Values of the $v_{\rm tot}$, $v_{\rm wt,st}$, and $K$ Parameters for the 14 Mini-BAL Quasars}
\tablehead{
    \colhead{} & \colhead{}                   & \colhead{$v_{\rm tot}$}  & \colhead{$v_{\rm wt,st}$}           
    & \colhead{$K$} \\
    \colhead{} & \colhead{Object Name (SDSS J)} & \colhead{(km~s$^{-1}$)}  & \colhead{(km~s$^{-1}$)}
    & \colhead{($10^{9}$ km$^3$~s$^{-3}$)}
}
\startdata
\multicolumn{2}{l}{{\it Chandra} Cycle 10 Objects} & &\\
& $080117.79+521034.5$ & $1095.8$ & $26641$ & $4331.9$ \\
& $091342.48+372603.3$ & $4958.8$ & $11695$ & $592.7$ \\ 
& $092914.49+282529.1$ & $1165.7$ & $24483$ & $4183.3$ \\
& $093207.46+365745.5$ & $1449.4$ & $3871$ & $21.9$ \\ 
& $105158.74+401736.7$ & $1104.4$ & $2383$ & $3.1$ \\
& $105904.68+121024.0$ & $8811.7$ & $6435$ & $813.3$ \\
& $120331.29+152254.7$ & $1098.5$ & $22118$ & $3067.5$ \\
& $123011.84+401442.9$ & $1242.5$ & $2326$ & $5.9$ \\
& $125230.84+142609.2$ & $1449.5$ & $2914$ & $9.4$ \\ 
& $141028.14+135950.2$ & $1725.7$ & $2394$ & $1.6$ \\
& $151451.77+311654.0$ & $2329.8$ & $26188$ & $5002.6$ \\ 
& $224649.29-004954.3$ & $1449.3$ & $4858$ & $53.6$ \\ \\
\multicolumn{2}{l}{Archival {\it XMM-Newton} Objects} & & \\
& $142656.17+602550.8$ & $1716.1$ & $27811$ & $4475.0$ \\ 
& $160222.72+084538.4$ & $2068.1$ & $14689$ & $487.2$ 
\enddata
\end{deluxetable}

The kinetic luminosity of an AGN outflow, $L_{\rm kin}\ (=\dot{M}v^2/2)$, is proportional to $N_Hv^3$, where 
$\dot{M}$ is the mass flux of the outflow, and $N_H$ is its column density (e.g., see \S 11.3 of Krolik 1999). We define 
another new parameter,
\begin{equation}
K=\frac{\int_{0}^{29,000}(1-f(v))\ v^3\ C^*dv}{v_{\rm tot}}.
\end{equation}
$K$ is a product of absorption strength, which is denoted by the factor $1-f(v)$, and $v^3$, averaged
over the velocity parameter space. It is 
heuristically motivated by kinetic luminosity, but it is not a proper measure of this quantity since, e.g., 
absorption strength is usually not proportional to $N_H$ (cf. Hamann 1998; Arav et al. 1999). For the high $S/N$ sample, 
the $K$ parameter is correlated with $\Delta\alpha_{\rm ox}$ at a probability of $97.1\%$ (see Fig.~11$c$). 
The values of $K$ for the 14 mini-BAL quasars in our sample are also listed in Table~8. 
 
For sources with a single broad absorption trough, $v_{\rm tot}$ and $v_{\rm wt,st}$ have the same values as
$\Delta v$ and $v_{\rm wt}$, respectively. For sources with multiple troughs, $v_{\rm tot}$ and $v_{\rm wt,st}$
have the advantage that they represent absorption features only. Furthermore, $v_{\rm tot}$ and $v_{\rm wt,st}$
have better correlations with the relative X-ray brightness. Although $v_{\rm tot}$ and $\Delta v$ have similar
high correlation probabilities with $\Delta\alpha_{\rm ox}$ ($>99.99\%$ vs. $99.94\%$), the figures 
(Fig.~11$a$ vs. Fig.~10$d$) show that the correlation between $v_{\rm tot}$ and $\Delta\alpha_{\rm ox}$ is stronger, 
which is also indicated by the Spearman rank-order correlation coefficients ($\rho=-0.682$ vs. $\rho=-0.463$).
We consider $v_{\rm tot}$ and $v_{\rm wt,st}$ to be more physically revealing parameters. The other newly 
introduced parameter $K$ does not have a better correlation with $\Delta\alpha_{\rm ox}$ than the simpler 
parameter $AI$ does. Further investigations may find additional useful parameters which could better 
represent the physical properties of UV absorbing outflows. 

\subsection{Investigation of a Very High Significance Mini-BAL Sample}
\begin{deluxetable}{rccc}
\tablecolumns{4} \tablenum{9} \tabletypesize{\footnotesize}
\tablewidth{\columnwidth}
\tablecaption{Spearman Rank-Order Tests for Very High Significance mini-BAL \& BAL Quasar Sample}
\tablehead{
     \colhead{} &  & \multicolumn{2}{c}{VHS mini-BAL + BAL Quasars}\\
     \colhead{} &  & \multicolumn{2}{c}{(500~km~s$^{-1}$)\tablenotemark{a}}
     \\
     \cline{3-4}\\
\colhead{} &  &  \colhead{$\rho$} & \colhead{$1-P_S$}
}
\startdata
                      $AI$ vs. $\Delta\ \alpha_{\rm ox}$ & & $-0.503$ & $99.92\%$\\	
              $v_{\rm max}$ vs. $\Delta\ \alpha_{\rm ox}$ & & $-0.427$ & $99.54\%$\\
              $v_{\rm min}$ vs. $\Delta\ \alpha_{\rm ox}$ & & $-0.157$ & $70.17\%$\\
              $\Delta v$ vs. $\Delta\ \alpha_{\rm ox}$ & & $-0.610$ & $99.99\%$\\
              $v_{\rm wt}$ vs. $\Delta\ \alpha_{\rm ox}$ & & $-0.270$ & $92.70\%$
\enddata
\tablenotetext{a}{Velocity width limit used when defining mini-BALs.}
\end{deluxetable}\label{table9}
In previous discussion and in G09, BAL and mini-BAL features were identified by formal criteria (Equations 1 \& 2).  
These formal criteria were used to distinguish mini-BALs from both narrower (NAL) and broader (BAL) absorption features. 
The identification process depends, to some 
extent, on continuum placement, spectral smoothing, and spectral $S/N$.  Because they are narrower and
(generally) weaker absorption 
features, mini-BALs can be more sensitive to small changes in these factors than BALs are. 
Mini-BALs selected according to the criteria in Equations 1 and 2 also exhibit a wide range of absorption 
trough morphologies.  The current work tests and expands upon the findings of G09 using spectra with absorption 
features that have been selected to represent robustly an intermediate class of absorption between BAL and NAL features.

As is apparent from Fig. 8, some formally selected mini-BAL troughs from G09 are not as significant as others. In order to 
investigate whether ``borderline" sources are affecting our results, we define a 
very high significance (VHS) mini-BAL quasar sample from the high $S/N$ sources in \S 4.4 by requiring the mini-BAL 
troughs lie at 
least 20\% below the continua, instead of 10\%. Under this stricter depth criterion and the width criterion of 
$500$~km~s$^{-1}$ (see \S 4.5 for discussion on the advantages of $500$~km~s$^{-1}$ over $1,000$~km~s$^{-1}$), 
24 of the 33 mini-BAL quasars remain in the VHS sample (see sources in 
Fig. 8 without asterisks following their names), including 13 of the 14 sources in our newly studied sample.
All the spectra of the VHS sample sources were examined by eye; they all show visually compelling intermediate absorption features.  

We repeated the analyses in \S4.2 and \S4.4 on relative \hbox{X-ray} brightness and its correlations with UV 
absorption properties for the VHS sample. The mean value of $\Delta\alpha_{\rm ox}$ is $-0.01\pm0.03$, 
which is consistent with that for mini-BAL quasars in \S4.2
($-0.03\pm0.02$). The correlation analyses were performed on the VHS mini-BAL quasars combined 
with the high $S/N$ BAL quasars in \S4.4. Two BAL quasars in G09 (SDSS~J0939+3556 and SDSS~J1205+0201) are 
removed here because they have $AI=0$ under
the 20\% depth criterion. The results are listed in Table~9. These correlations for VHS mini-BAL and BAL 
quasars are generally consistent with those for the high $S/N$ sample in \S4.4 (see Table~6). In summary, the factors
in selection of mini-BALs (e.g., the placement of continua) do not significantly affect the main results
of our analyses.  
 

\section{Conclusions}

We present the \hbox{X-ray} properties of 14 of the optically brightest mini-BAL quasars from SDSS DR5; 12
have been observed in a new {\it Chandra} snapshot survey in Cycle 10 and 2 objects
have archival {\it XMM-Newton} observations. All 14 sources are detected. We study 
correlations between the UV absorption properties and the \hbox{X-ray} brightness for a sample consisting
of these 14 sources, as well as the BAL and mini-BAL quasars in G09. Our main results are summarized 
as follows:
\begin{enumerate}
\item The mean \hbox{X-ray} power-law photon index of the 12 {\it Chandra} observed sources is 
$\Gamma=1.90^{+0.11}_{-0.11}$ for a model without intrinsic absorption. Adding an intrinsic neutral 
absorption component only slightly changes the mean photon index to $\Gamma=1.91^{+0.18}_{-0.11}$, 
consistent with the absence of evidence for strong intrinsic neutral absorption on average
($N_H \lesssim 8.2\times 10^{21} \ {\rm cm}^{-2}$). This indicates that mini-BAL quasars generally have similar
\hbox{X-ray} spectral properties to non-BMB quasars. 
\item The relative \hbox{X-ray} brightness, assessed with the $\Delta\alpha_{\rm ox}$ parameter, of the 
mini-BAL quasars has a closer mean value and distribution to those of non-BMB quasars than to
those of BAL quasars. An Anderson-Darling test finds no non-normality of the $\Delta\alpha_{\rm ox}$
distribution of \hbox{mini-BAL} quasars.
\item The reddening, assessed with the parameter $R_{\rm red}$, of the mini-BAL quasars is intermediate between  
those of non-BMB and HiBAL quasars. Relative colors show curvature in the optical spectra of mini-BAL quasars, 
indicating mild dust reddening. 
\item Significant correlations are found between $\Delta\alpha_{\rm ox}$ and $AI$, $v_{\rm max}$ and 
$\Delta v$. The weighted average velocity $v_{\rm wt}$ has a marginal correlation with 
$\Delta\alpha_{\rm ox}$. These correlations
 may support the radiatively driven disk-wind model where \hbox{X-ray} absorption is important in enabling 
UV line-absorbing winds. 
\item We find that more intrinsic broad absorption troughs are included if the velocity width 
limit in the definition of mini-BALs is lowered to $500$~km~s$^{-1}$. The UV absorption parameters
$AI$, $v_{\rm max}$, $v_{\rm min}$, $\Delta v$, and $v_{\rm wt}$ have clearer correlations with the relative 
\hbox{X-ray} brightness under the new velocity width limit. We also propose three new parameters, $v_{\rm tot}$, 
$v_{\rm wt, st}$, and $K$, defined by Equations (9), (10), and (11), respectively. $v_{\rm tot}$ is the 
total velocity span of all the UV broad
absorption troughs, while $v_{\rm wt, st}$ is the weighted average velocity for the strongest absorption trough. 
We consider $v_{\rm tot}$ and $v_{\rm wt,st}$ to represent better the physical properties of UV absorption features
than $\Delta v$ and $v_{\rm wt}$.
$K$ is similar to, but not a proper measure of, the kinetic luminosity of the absorbing outflows. 
\item Testing shows that the complex factors in the selection of mini-BAL quasars, such as the placement of continua, 
do not significantly affect 
our main results on the relative X-ray brightness and its correlations with UV absorption properties. 
\end{enumerate}

Further accumulation of high-quality \hbox{X-ray} and UV/optical 
data will allow studies of even larger samples of mini-BAL and 
BAL quasars. The improved source statistics should enable the
construction and testing of further measures of mini-BAL and 
BAL absorption that may provide additional insight in the 
exploration of the broad absorption region. Ultimately, 
high-resolution \hbox{X-ray} spectroscopy performed by future 
missions, such as $IXO$, should reveal the detailed physical
and kinematic properties of the \hbox{X-ray} absorber. This 
will open a new dimension in studies of \hbox{X-ray} vs.\,UV/optical 
absorption for quasar outflows. 


\acknowledgments{
We thank the anonymous referee for constructive comments. 
We gratefully acknowledge the financial support of NASA grant SAO 
SV4-74018 (G.P.G., Principal Investigator), NASA ADP grant NNX10AC99G 
(J.W., W.N.B., M.L.C.), NASA Chandra grant AR9-0015X (R.R.G.), 
and NSF grant AST07-09394 (R.R.G.). We thank Y. Q. Xue for helpful discussions.

Funding for the SDSS and SDSS-II has been provided by the Alfred P. Sloan Foundation, 
the Participating Institutions, the National Science Foundation, the U.S. Department 
of Energy, the National Aeronautics and Space Administration, the Japanese Monbukagakusho, 
the Max Planck Society, and the Higher Education Funding Council for England. The SDSS Web 
Site is http://www.sdss.org/.
}





\clearpage
\begin{turnpage}
\begin{deluxetable}{lcccccccccccccc}
\tablecolumns{15} \tablenum{4} \tabletypesize{\scriptsize}
\tablewidth{0pt}
\tablecaption{X-Ray, Optical, and Radio Properties}
\tablehead{
    \colhead{} 
    & \colhead{}                     
    & \colhead{}    
    & \colhead{}        
    & \colhead{}        
    & \colhead{}            
    & \colhead{Count}
    & \colhead{}                       
    & \colhead{}                 
    & \colhead{log $L$}                 
    & \colhead{}
    & \colhead{log $L_{\nu}$}          
    & \colhead{} 
    & \colhead{} 
    & \colhead{} \\
    \colhead{} 
    & \colhead{Object Name (SDSS J)} 
    & \colhead{$z$} 
    & \colhead{$m_{i}$\tablenotemark{a}} 
    & \colhead{$M_{i}$} 
    & \colhead{$N_{\rm H}$} 
    & \colhead{Rate\tablenotemark{b}} 
    & \colhead{$F_{0.5-2\;{\rm keV}}$\tablenotemark{c}} 
    & \colhead{$f_{\rm 2\;keV}$\tablenotemark{d}} 
    & \colhead{($2-10\;{\rm keV}$)} 
    & \colhead{$f_{2500\mbox{\rm~\scriptsize\AA}}$\tablenotemark{e}} 
    & \colhead{(2500 \AA)}  
    & \colhead{$\alpha_{\rm ox}$} 
    &  \colhead{$\Delta \alpha_{\rm ox}$ $(\sigma)$\tablenotemark{f}} 
    & \colhead{$R$} \\
    \colhead{} 
    & \colhead{(1)} 
    & \colhead{(2)} 
    & \colhead{(3)} 
    & \colhead{(4)} 
    & \colhead{(5)} 
    & \colhead{(6)} 
    & \colhead{(7)} 
    & \colhead{(8)} 
    & \colhead{(9)} 
    & \colhead{(10)} 
    & \colhead{(11)} 
    & \colhead{(12)} 
    & \colhead{(13)} 
    & \colhead{(14)}
}
\startdata
\multicolumn{2}{l}{{\it Chandra} Cycle 10 Objects} & & & & & & & & & & \\
& $080117.79+521034.5$ & $3.23$ & $16.76$ & $-29.78$ & $4.58$ 
& $4.92 ^{+1.35}_{-1.09}$ &  $1.99$    & $12.52$ & $45.33$  
& $8.17$  & $32.25$ &  $-1.85$ &  $-0.04$ $(0.29)$ & $<0.51$ \\
& $091342.48+372603.3$ & $2.13$ & $17.38$ & $-28.16$ & $1.91$ 
& $7.67 ^{+1.44}_{-1.23}$ & $2.87$    & $13.40$ & $45.05$  
& $3.60$  & $31.59$ &  $-1.70$ &  $ 0.02$ $(0.12)$ & $<0.73$ \\
& $092914.49+282529.1$ & $3.41$ & $17.45$ & $-29.14$ & $2.06$ 
& $6.02 ^{+1.31}_{-1.09}$ &  $2.26$    & $ 14.84$ & $45.44$  
& $4.28$  & $32.01$ &  $-1.71$ &  $0.07$ $(0.49)$ & $<0.74$ \\
& $093207.46+365745.5$ & $2.90$ & $17.42$ & $-28.80$ & $1.37$ 
& $9.85 ^{+1.54}_{-1.34}$ &  $3.63$    & $21.10$ & $45.48$  
& $3.98$  & $31.86$ &  $-1.64$ &  $ 0.11$ $(0.78)$ & $<1.24$ \\
& $105158.74+401736.7$ & $2.17$ & $16.71$ & $-28.87$ & $1.33$ 
& $41.22^{+3.43}_{-3.17}$ &  $15.16$   & $71.67$ & $45.79$  
& $6.75$  & $31.88$ &  $-1.53$ &  $ 0.23$ $(1.60)$ & $<0.31$ \\
& $105904.68+121024.0$ & $2.50$ & $17.43$ & $-28.47$ & $2.11$ 
& $8.45 ^{+1.69}_{-1.42}$ &  $3.18$    & $16.62$ & $45.27$  
& $3.96$  & $31.75$ &  $-1.68$ &  $ 0.06$ $(0.41)$ & $<0.68$ \\
& $120331.29+152254.7$ & $2.99$ & $16.87$ & $-29.45$ & $3.03$ 
& $18.82^{+2.41}_{-2.15}$ &  $7.27$    & $43.09$ & $45.81$  
& $7.28$  & $32.14$ &  $-1.62$ &  $ 0.17$ $(1.31)$ & $<0.57$ \\
& $123011.84+401442.9$ & $2.05$ & $16.97$ & $-28.48$ & $1.66$ 
& $12.09^{+1.76}_{-1.55}$ &  $4.50$    & $20.44$ & $45.21$  
& $5.11$  & $31.71$ &  $-1.69$ &  $ 0.05$ $(0.32)$ & $<0.54$ \\
& $125230.84+142609.2$ & $1.94$ & $16.55$ & $-28.81$ & $2.13$ 
& $10.05^{+1.86}_{-1.58}$ &  $3.80$    & $16.69$ & $45.07$  
& $7.37$  & $31.83$ &  $-1.78$ &  $-0.03$ $(0.22)$ & $<0.49$ \\
& $141028.14+135950.2$ & $2.22$ & $17.30$ & $-28.34$ & $1.39$ 
& $12.41^{+2.00}_{-1.74}$ &  $4.57$    & $21.89$ & $45.29$  
& $4.21$  & $31.68$ &  $-1.64$ &  $ 0.09$ $(0.59)$ & $<0.79$ \\
& $151451.77+311654.0$ & $2.14$ & $17.06$ & $-28.49$ & $1.84$ 
& $4.78 ^{+1.34}_{-1.07}$ &  $1.79$    & $8.33$  & $44.84$  
& $4.88$  & $31.72$ &  $-1.83$ &  $-0.09$ $(0.64)$ & $<0.26$ \\
& $224649.29-004954.3$ & $2.04$ & $17.47$ & $-28.10$ & $5.13$ 
& $11.37^{+1.70}_{-1.49}$ &  $4.67$    & $21.18$ & $45.22$  
& $3.59$  & $31.55$ &  $-1.62$ &  $ 0.09$ $(0.61)$ & $<1.28$ \\ \\
\multicolumn{2}{l}{Archival {\it XMM-Newton} Objects} & & & & & & & & & & \\
& $142656.17+602550.8$ & $3.20$ & $16.21$ & $-30.22$ & $1.75$ 
& $23.86^{+1.99}_{-1.84}$ &  $6.02$    & $37.72$ & $45.80$ 
& $12.48$ & $32.43$ &  $-1.82$ &  $ 0.02$ $(0.12)$ & $<0.50$ \\
& $160222.72+084538.4$ & $2.28$ & $17.05$ & $-28.69$ & $3.71$ 
& $11.68^{+1.92}_{-1.66}$\tablenotemark{g} &  $3.18$    & $15.48$ & $45.16$ 
& $5.58$  & $31.83$ &  $-1.75$ &  $ 0.00$ $(0.02)$ & $<0.99$ 
\enddata 
\tablenotetext{a}{The apparent $i$-band magnitude using the SDSS quasar catalog BEST 
photometry.}
\tablenotetext{b}{The count rate in the observed-frame soft \hbox{X-ray} 
band ($0.5-2.0$ keV) in units of $10^{-3}$ ${\rm s}^{-1}$.}
\tablenotetext{c}{The Galactic absorption-corrected observed-frame flux between 
$0.5-2.0$ keV in units of $10^{-14}$ ergs cm$^{-2}$ s$^{-1}$.} 
\tablenotetext{d}{The flux density at rest-frame 2 keV, in units of 
$10^{-32}$ ergs cm$^{-2}$ s$^{-1}$ Hz$^{-1}$.}
\tablenotetext{e}{The flux density at rest-frame 2500~\AA\ in units of 
10$^{-27}$ ergs cm$^{-2}$ s$^{-1}$ Hz$^{-1}$.}
\tablenotetext{f}{$\Delta\alpha_{\rm ox}$: the difference between the 
measured $\alpha_{\rm ox}$ and the expected $\alpha_{\rm ox}$, 
defined by the $\alpha_{\rm ox}-L_{2500~{\rm \AA}}$ 
relation in equation (3) of Just et al.~(2007); 
the statistical significance of this difference, $\sigma$, 
is measured in units of the RMS $\alpha_{\rm ox}$ defined in Table 5 of 
Steffen et al.~(2006).}
\tablenotetext{g}{Calculated using an effective exposure time corrected for vignetting
at large off-axis angle.}
\label{table4}
\end{deluxetable}
\clearpage
\end{turnpage}

\begin{figure*}[h!tbp]
    \centering
    \includegraphics[width=3.2in]{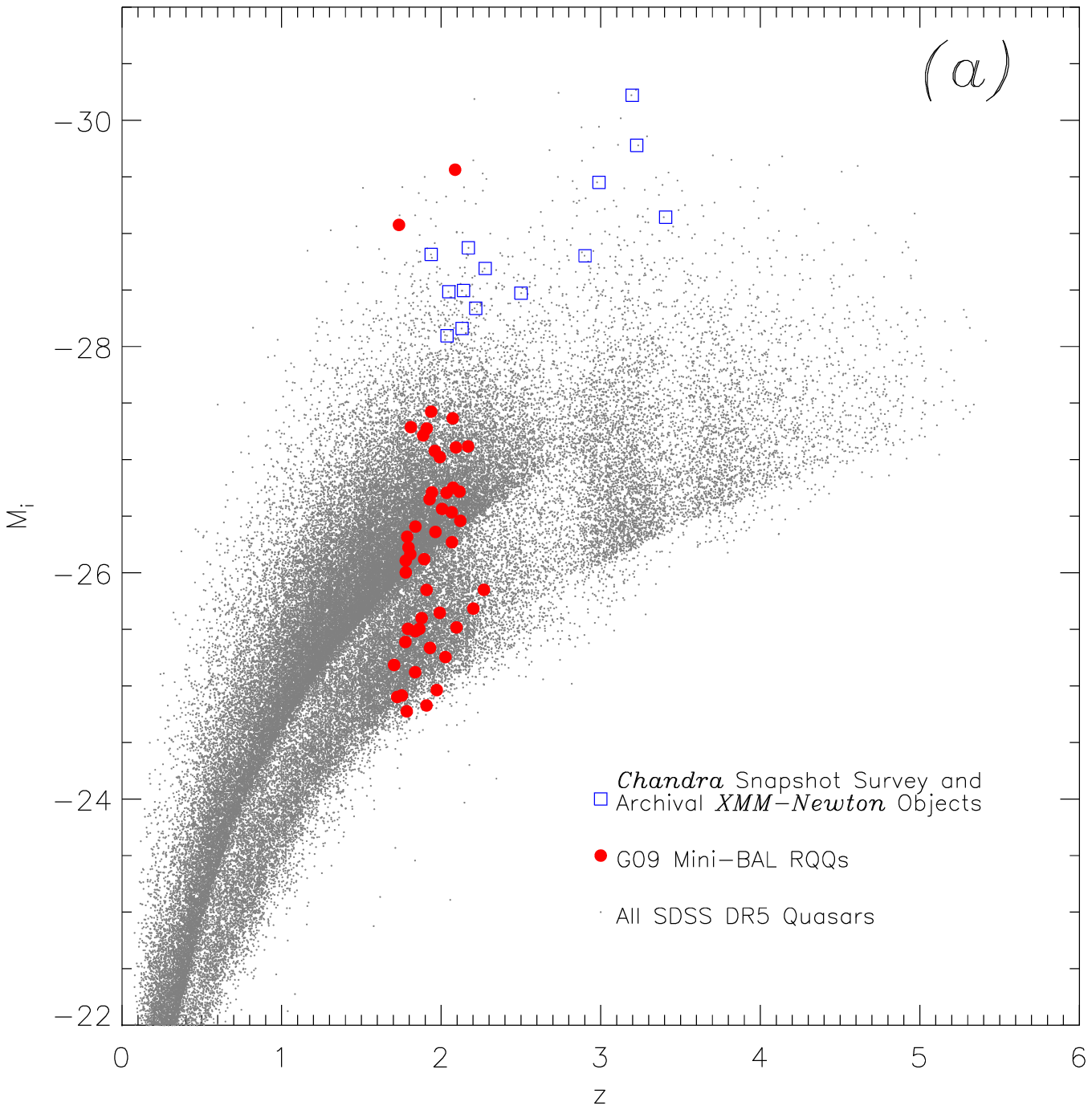}
    \includegraphics[width=3.2in]{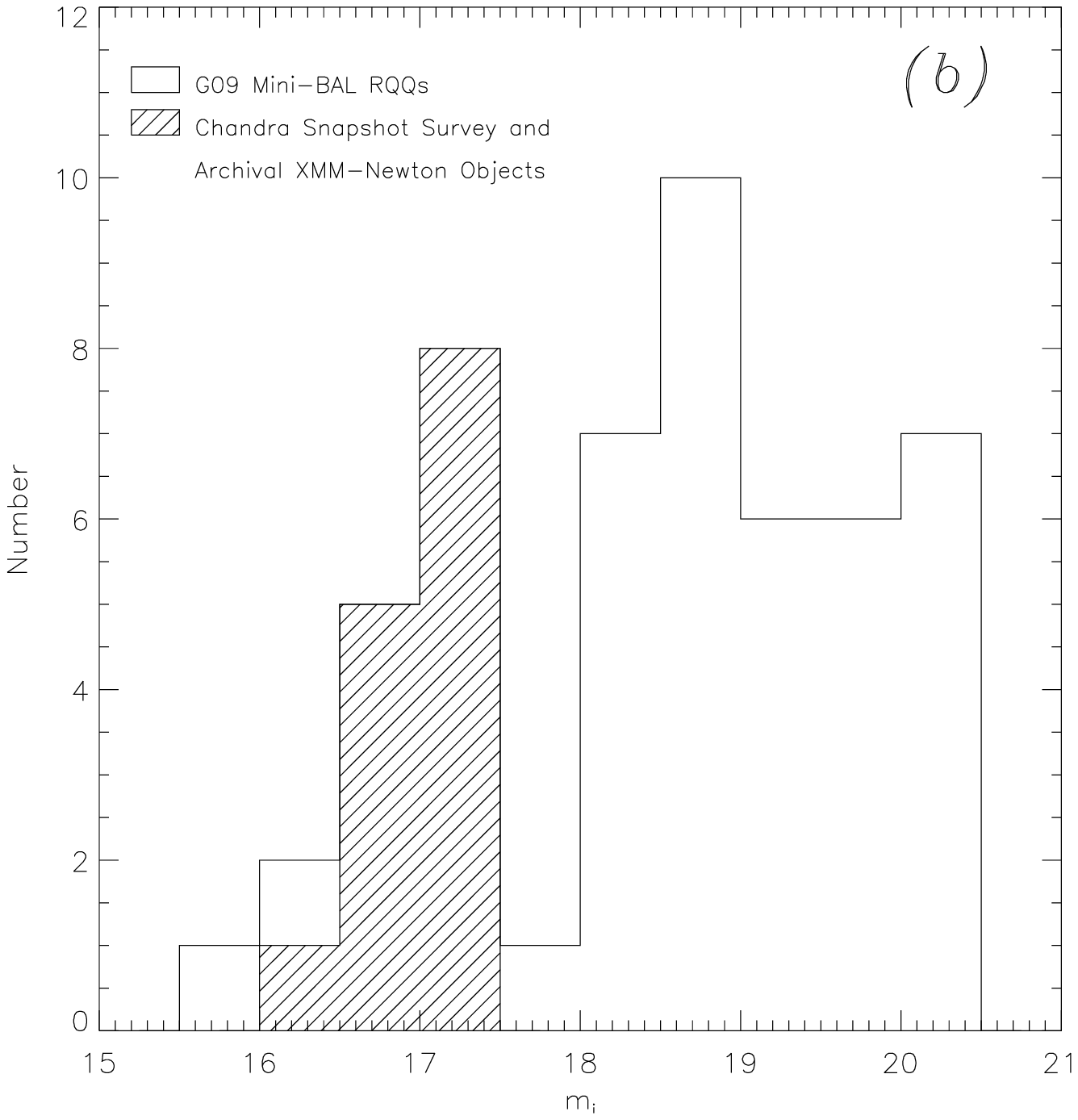}
    \includegraphics[width=3.2in]{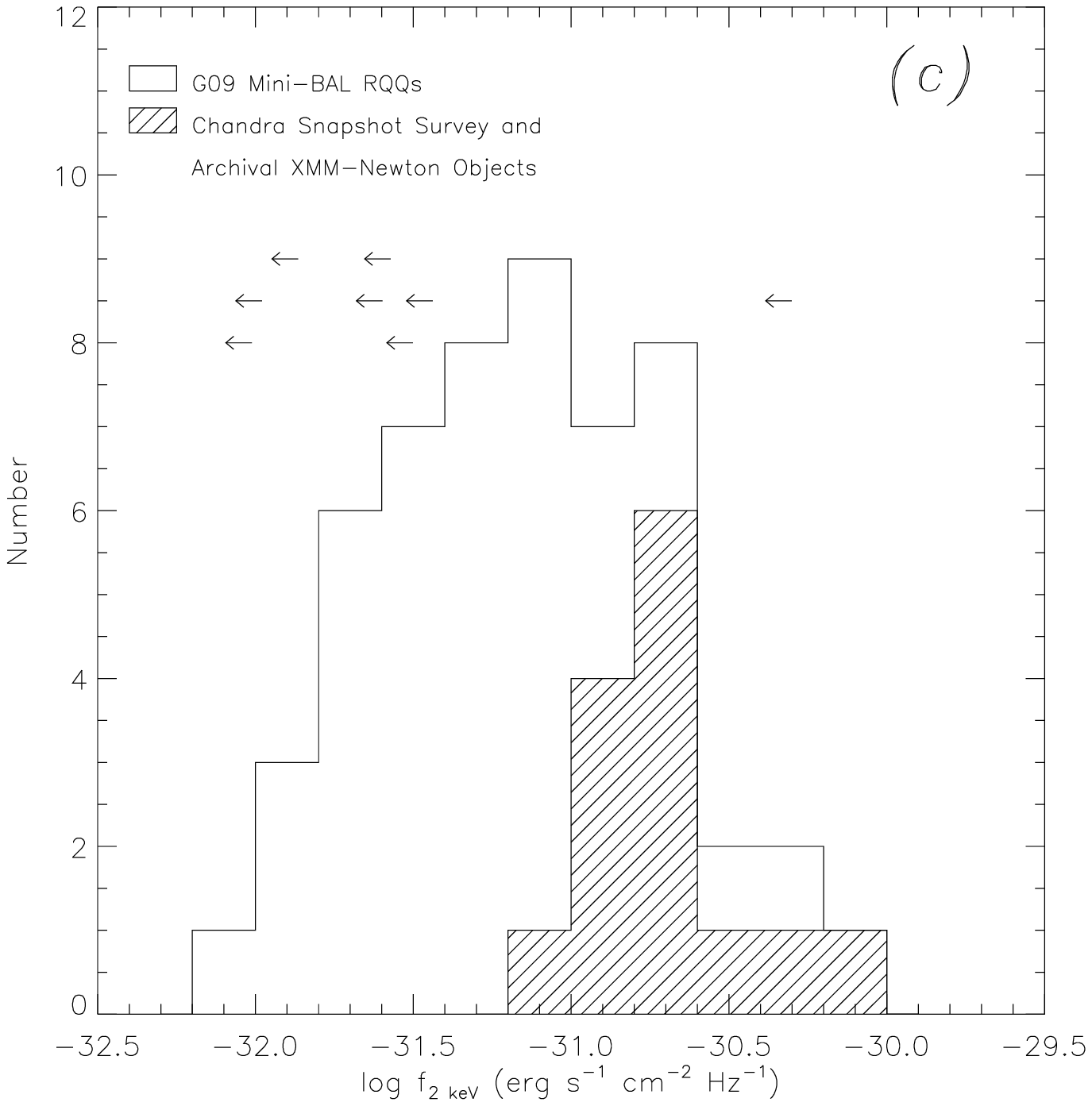}
    \caption{\footnotesize{{\it (a)} SDSS absolute $i$-band magnitude, $M_{i}$, plotted with 
             respect to redshift, $z$. The blue squares represent the 14 
             mini-BAL quasars in our combined {\it Chandra} snapshot survey 
             and archival {\it XMM-Newton} mini-BAL quasar sample, the red 
             circles represent the 48 mini-BAL quasars in G09,
             and the small black dots represent the 77,429 objects in the 
             SDSS DR5 quasar catalog.
             {\it (b)} Histogram showing the distribution of SDSS $i$-band 
             apparent magnitude, $m_{i}$, for the objects observed in our 
             combined {\it Chandra} snapshot survey and archival 
             {\it XMM-Newton} mini-BAL quasar sample, relative to the 48 
             radio-quiet mini-BAL quasars in G09.
             {\it (c)} Histogram showing the distribution of \hbox{X-ray} flux 
             density at rest-frame 2~keV for the objects observed in our 
             combined {\it Chandra} snapshot survey and archival 
             {\it XMM-Newton} mini-BAL quasar sample, relative to the 48 
             radio-quiet mini-BAL quasars in G09. Leftward arrows show undetected 
             G09 sources, with arbitrary $y$-coordinates.}}
             \label{fig1}
\end{figure*}

\begin{figure*}[t]
    \centering
    \includegraphics[width=6.5in]{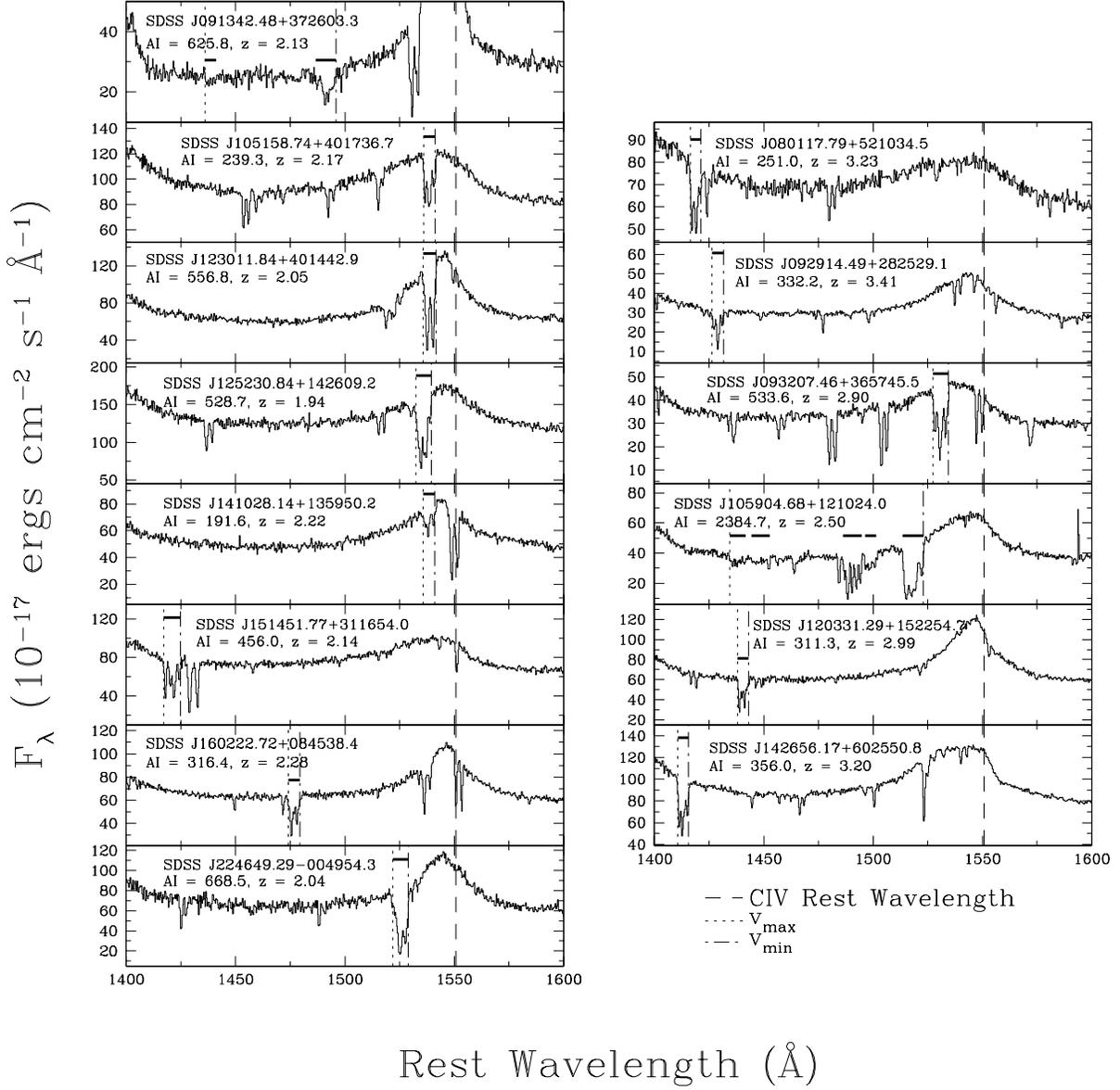}
    \caption{\footnotesize{SDSS spectra for our {\it Chandra} snapshot survey sample, as 
             well as the included archival {\it XMM-Newton} \hbox{mini-BAL} quasar 
             observations, plotted with respect to rest-frame wavelength in 
             Angstroms, using the improved redshift measurements in Hewett \& Wild (2010). 
             The region(s) contributing to the absorption index, 
             $AI$, for the C~{\sc iv}~$\lambda$1549 mini-BAL are marked by 
             horizontal lines over the relevant trough(s). The $y$-coordinates 
             of these horizontal lines are arbitrary. On each spectrum, the 
             C~{\sc iv} rest wavelength is marked by a dashed line at 
             1550.77${\rm \AA}$; the maximum outflow velocity, $v_{\rm max}$, 
             is marked by a dotted line; and the minimum outflow velocity, 
             $v_{\rm min}$, is marked by a dot-dashed line. Each panel is 
             labeled with the SDSS object name, $AI$, and redshift. The 
             left column corresponds to quasars at 1.9 $\leq\;z\;<$ 2.5, and 
             the right column corresponds to quasars at 2.5 $\leq\;z\;<$ 3.5. 
             Dividing the sources into two columns is only for presentation purposes.  
             Each spectrum is corrected for Galactic extinction, according to 
             Cardelli et al.~(2009), and for fiber light loss, according to
             the procedure outlined in \S 2. The spectral resolution ($\lambda/\Delta\lambda$)
             is $\approx$~2000.}}
             \label{fig2}
\end{figure*}

\begin{figure*}[h!tbp]
    \centering
    \includegraphics[angle=0, width=6.5in]{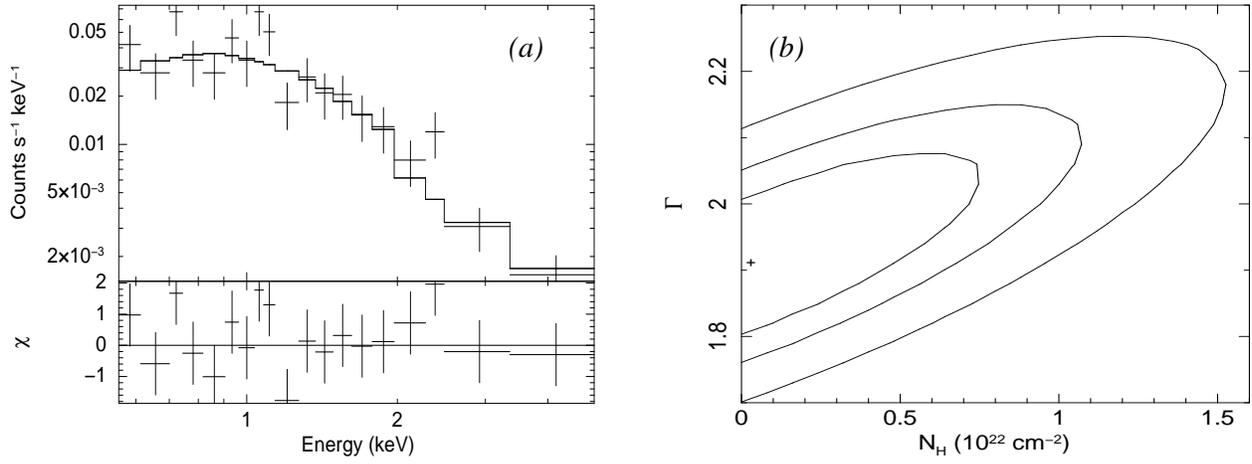}
    \caption{\footnotesize{{\it (a)} \hbox{X-ray} spectrum of SDSS J1051+4017 (binned to at least 10 counts
              per bin for the purpose of presentation) fitted with the power-law model
              with both Galactic and intrinsic absorption. The residuals are
              shown in units of $\sigma$. 
             {\it (b)} The contour map of the photon index vs. intrinsic column density
              at confidence levels of 68\%, 90\% and 99\%, for the joint fitting of 12 
              {\it Chandra} snapshot survey sources.}} 
             \label{fig3}
\end{figure*}

\begin{figure}[htbp]
    \centering
    \includegraphics[width=5.0in]{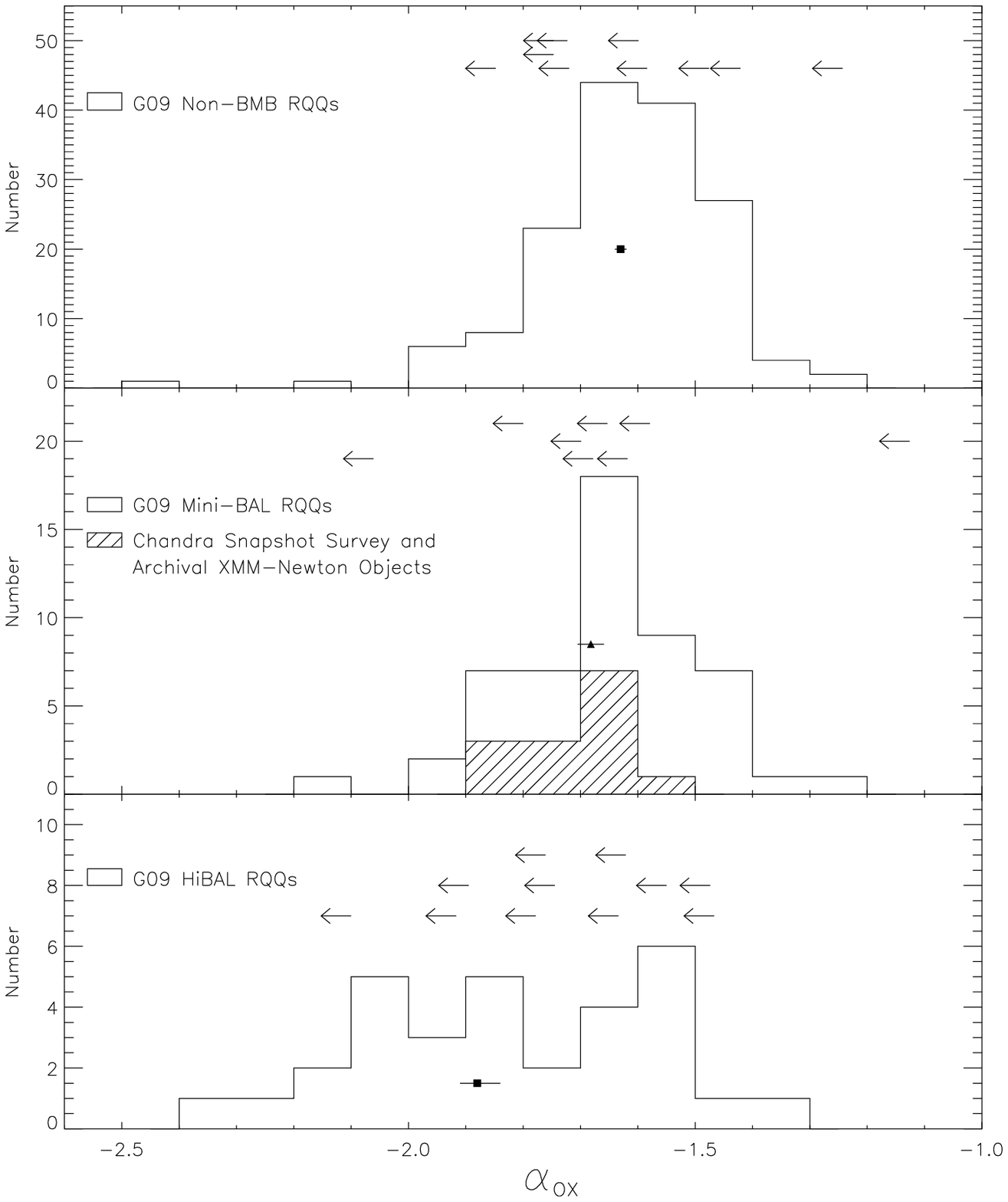}
    \caption{\footnotesize{{\it Top panel}: Histogram showing the distribution of 
              $\alpha_{\rm ox}$ for the radio-quiet non-BMB quasars in G09. 
              Upper limits corresponding to 1$\sigma$ for the 
              non-detections are indicated by arrows, where the $y$-coordinates
              of these arrows are arbitrary. The mean value of 
              $\alpha_{\rm ox}$, described in \S 4.2, for the radio-quiet 
              non-BMB quasars in G09 is shown by a filled square and 
              its horizontal error bars. The $y$-coordinate of the mean-value 
              point is arbitrary, and the error bars are so small that they 
              are only slightly visible on the plot.
              {\it Middle panel}: Histogram showing the distribution of 
              $\alpha_{\rm ox}$ for the objects observed in our {\it Chandra} 
              snapshot survey and archival {\it XMM-Newton} mini-BAL quasar 
              sample, relative to the 48 radio-quiet mini-BAL quasars from 
              G09. Upper limits corresponding to 1$\sigma$ 
              for the non-detected mini-BAL quasars in G09 
              are indicated by arrows. The mean value of $\alpha_{\rm ox}$ for 
              our mini-BAL quasar sample, combined with the mini-BAL quasars 
              in G09, is shown by a filled triangle and 
              its horizontal error bars.
              {\it Bottom panel}: Histogram showing the distribution of 
              $\alpha_{\rm ox}$ for the radio-quiet HiBAL quasars presented in 
              G09. Upper limits corresponding to 1$\sigma$ for 
              non-detections are indicated by arrows. The mean value of 
              $\alpha_{\rm ox}$ for the HiBAL quasars in G09 
              is shown by a filled square and its horizontal error bars.}} 
              \label{fig4}
\end{figure}

\begin{figure}[htbp]
    \centering
    \includegraphics[width=5.0in]{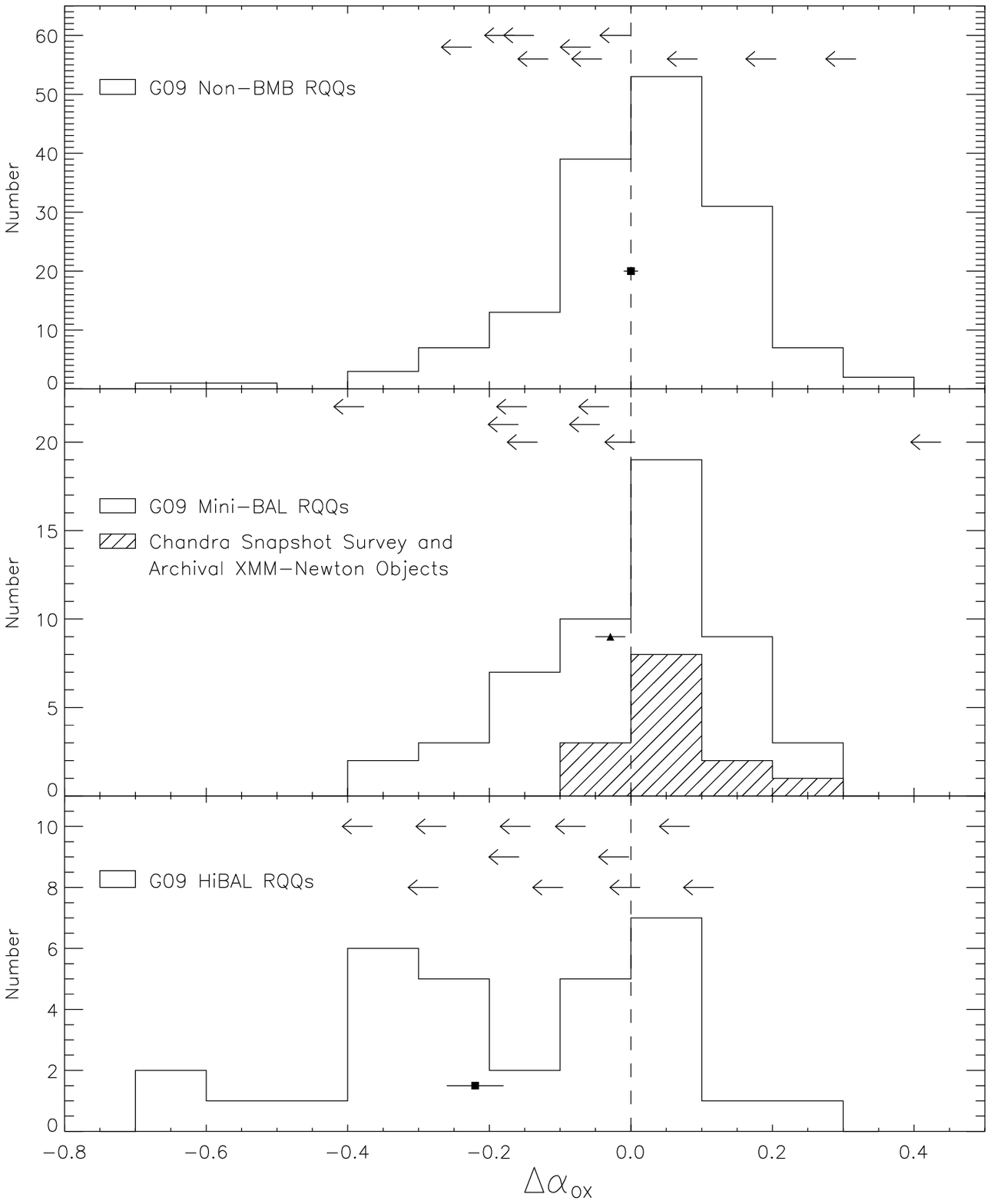}
    \caption{\footnotesize{{\it Top panel}: Histogram showing the distribution of 
             $\Delta\alpha_{\rm ox}$ for the radio-quiet non-BMB quasars in 
             G09. Upper limits corresponding to 1$\sigma$ 
             for non-detections are indicated by arrows, where the 
             $y$-coordinates of these arrows are arbitrary. 
             The mean value of $\Delta\alpha_{\rm ox}$, described in \S 4.2, 
             for the non-BMB quasars in G09 is shown by a 
             filled square and its horizontal error bars. The $y$-coordinate 
             of the mean-value point is arbitrary, and the error bars are so 
             small that they are only slightly visible on the plot.
             {\it Middle panel}: Histogram showing the distribution of 
             $\Delta\alpha_{\rm ox}$ for the objects observed in our combined 
             {\it Chandra} snapshot survey and archival {\it XMM-Newton} 
             mini-BAL quasar sample, relative to the 48 radio-quiet mini-BAL 
             quasars from G09. Upper limits corresponding 
             to 1$\sigma$ for the non-detected mini-BAL quasars in G09 
             are indicated by arrows. The mean value of 
             $\Delta\alpha_{\rm ox}$ for our mini-BAL quasar sample, combined 
             with the mini-BAL quasars in G09, 
             is shown by a filled triangle and its horizontal error bars. 
             {\it Bottom panel}: Histogram showing the distribution of 
             $\Delta\alpha_{\rm ox}$ for the radio-quiet HiBAL quasars presented 
             in G09. Upper limits corresponding to 1$\sigma$ 
             for non-detections are indicated by arrows. The mean value of 
             $\Delta\alpha_{\rm ox}$ for the HiBAL quasars in G09 is shown by a 
             filled square and its horizontal error bars. 
             In all three panels, the dashed line indicates 
             $\Delta\alpha_{\rm ox} = 0$.}}
             \label{fig5}
\end{figure}

\begin{figure}[htbp]
    \centering
    \includegraphics[width=5.8in]{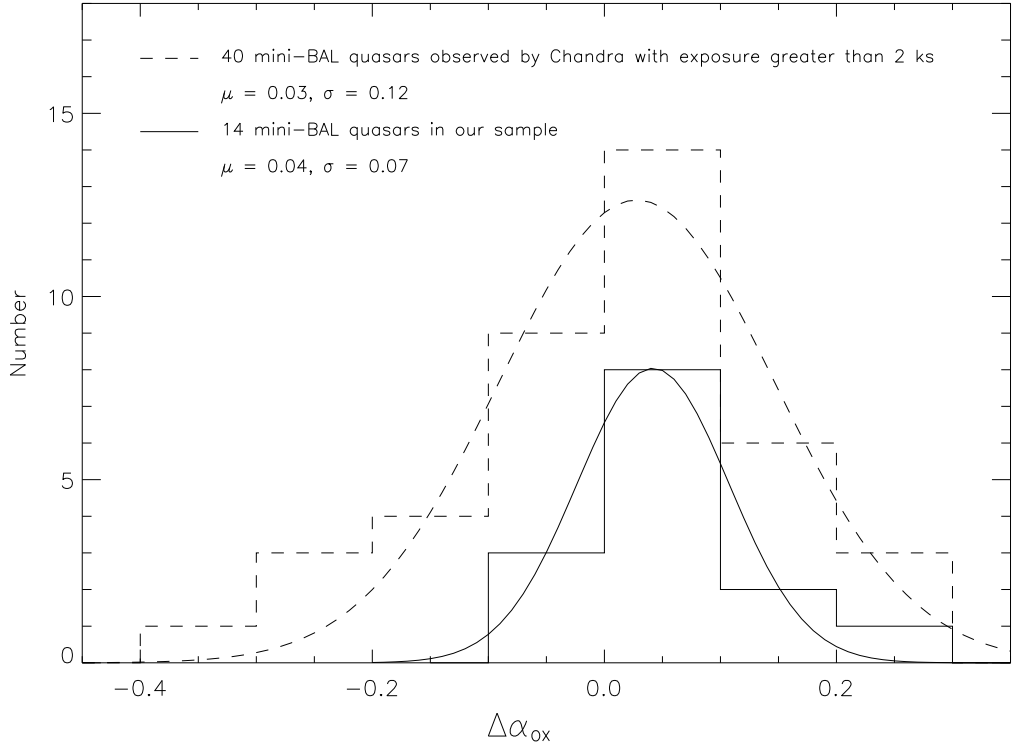}
    \caption{\footnotesize{Normality tests of the $\Delta\alpha_{\rm ox}$ values for mini-BAL 
             quasars. The bold histogram and curve represent the distribution of 
             $\Delta\alpha_{\rm ox}$ for the 14 sources in our sample and its 
             best-fit Gaussian. The dotted histogram and curve represent 
             the distribution of $\Delta\alpha_{\rm ox}$ and its best-fit Gaussian 
             for the 31 sources observed by {\it Chandra} with an exposure time 
             greater than 2~ks.}}
             \label{fig6}
\end{figure}

\begin{figure}[htbp]
    \centering
    \includegraphics[width=5.8in]{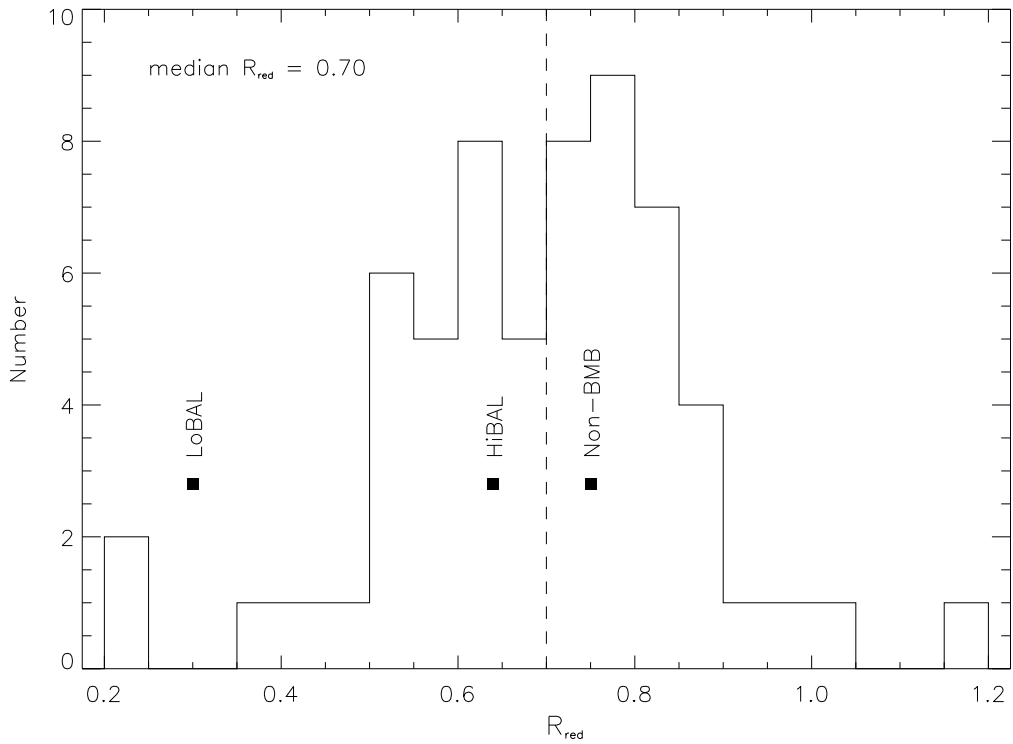}
    \caption{\footnotesize{Distribution of the ratio $R_{\rm red}\equiv F_{\nu}(1400 \ {\rm \AA})/F_{\nu}(2500 \ {\rm \AA})$
             for the 61 mini-BAL quasars in the combined sample. The vertical dashed line indicates the median
             $R_{\rm red}$ value (0.70) for mini-BAL quasars. The three filled squares show the median 
             $R_{\rm red}$ values for non-BMB quasars (0.75), HiBAL quasars (0.64), and LoBAL quasars (0.30)
             in Gibson et al. (2009b), respectively.}}
             \label{fig7}
\end{figure}

\begin{figure}[htbp]
    \centering
    \includegraphics[width=5.5in]{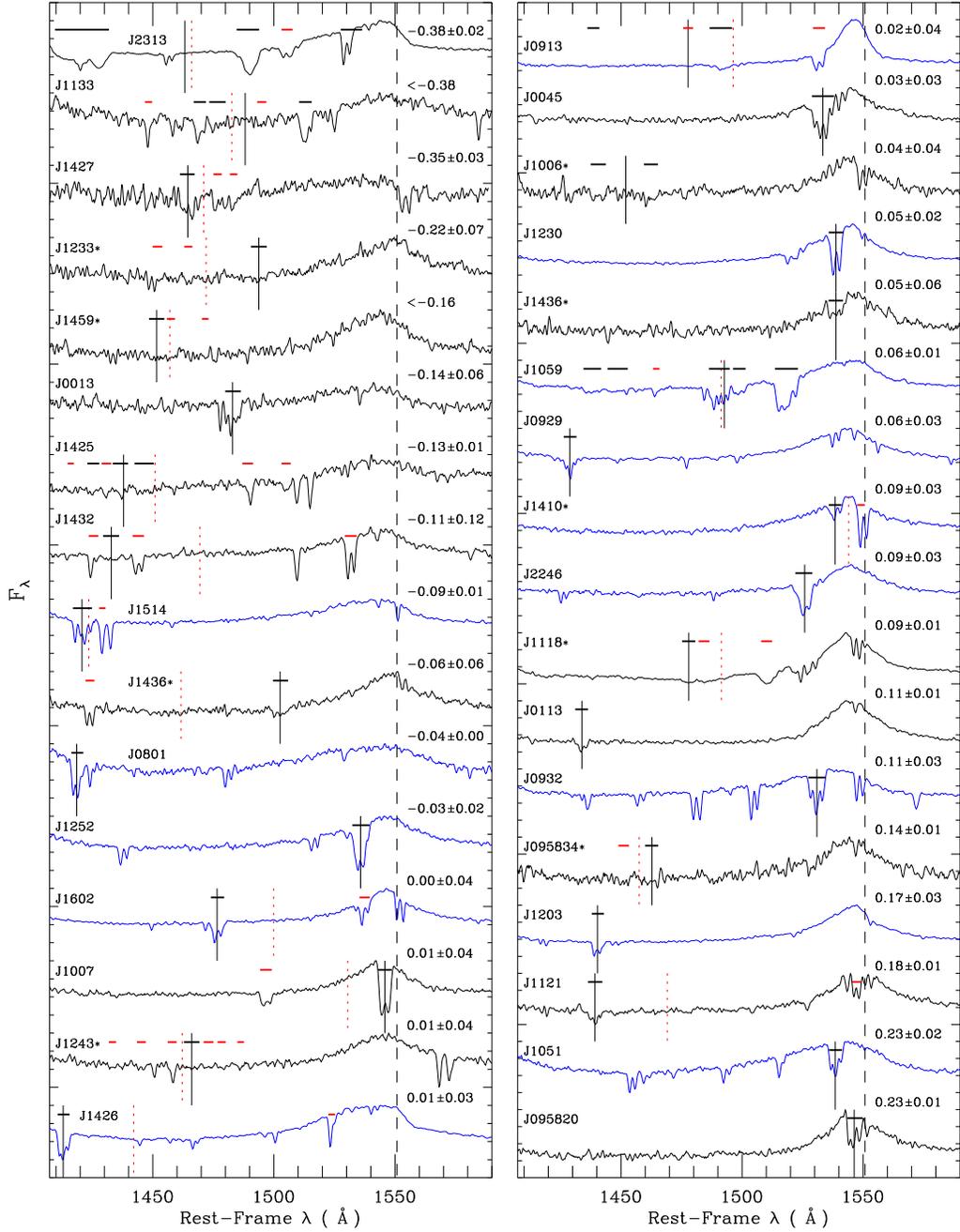}
    \caption{\footnotesize{SDSS spectra for the 14 mini-BAL quasars in our sample (blue curves) and 
             the high $S/N$ mini-BAL quasars in G09 (black curves), in the order of 
             $\Delta\alpha_{\rm ox}$. The $\Delta\alpha_{\rm ox}$ values and their 
             error bars (if the source is detected in \hbox{X-rays}) are shown for each source. 
             The name of each source is labeled in the format of 'J{\it hhmm}' for brevity (except J095820 \&
             J095834).
             The $y$-coordinates are arbitrary. The left threshold of the $x$-coordinates (1407.36~\AA) corresponds 
             to the velocity integration limit of $29,000$~km~s$^{-1}$ in the definition of mini-BALs. 
             As in Fig.~2, the dashed lines show the rest-frame wavelength (1550.77~\AA) 
             of C~{\sc iv}. For each spectrum, the black horizontal bars and the black solid vertical line 
             show the absorption troughs and weighted average velocity $v_{\rm wt}$ (see Equation (7) for definition) 
             under the velocity width limit of $1,000$~km~s$^{-1}$, 
             respectively. The red horizontal bars show the absorption troughs that would
             be included if the velocity width limit were chosen to be $500$~km~s$^{-1}$. The red dotted 
             vertical line shows the accordingly changed $v_{\rm wt}$. Sources with asterisks after their names are
             excluded in the very high significance mini-BAL quasar sample as defined in \S4.6.}}
             \label{fig8}
\end{figure}

\begin{figure}[htbp]
    \centering
    \includegraphics[width=2.4in]{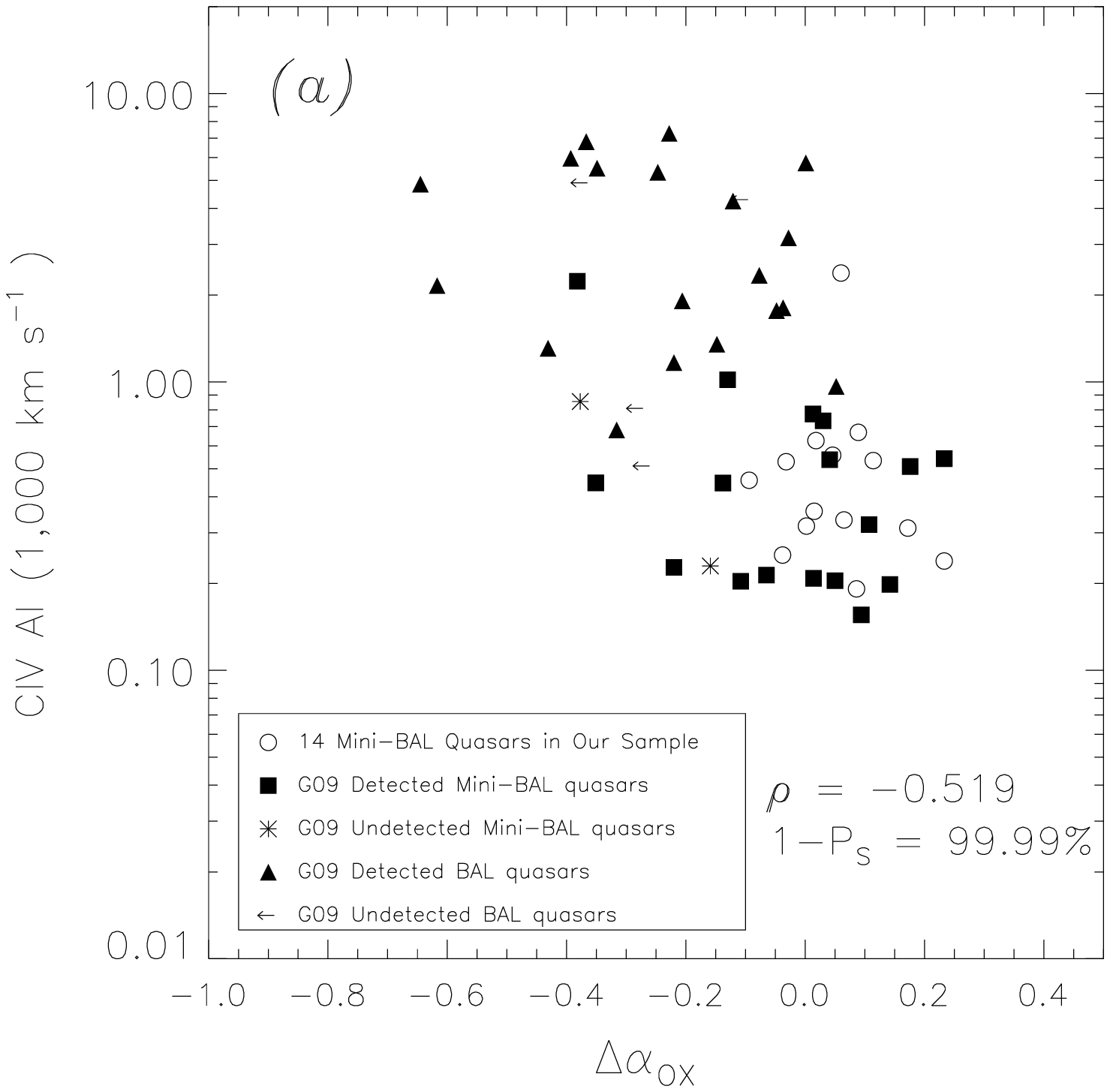}
    \includegraphics[width=2.4in]{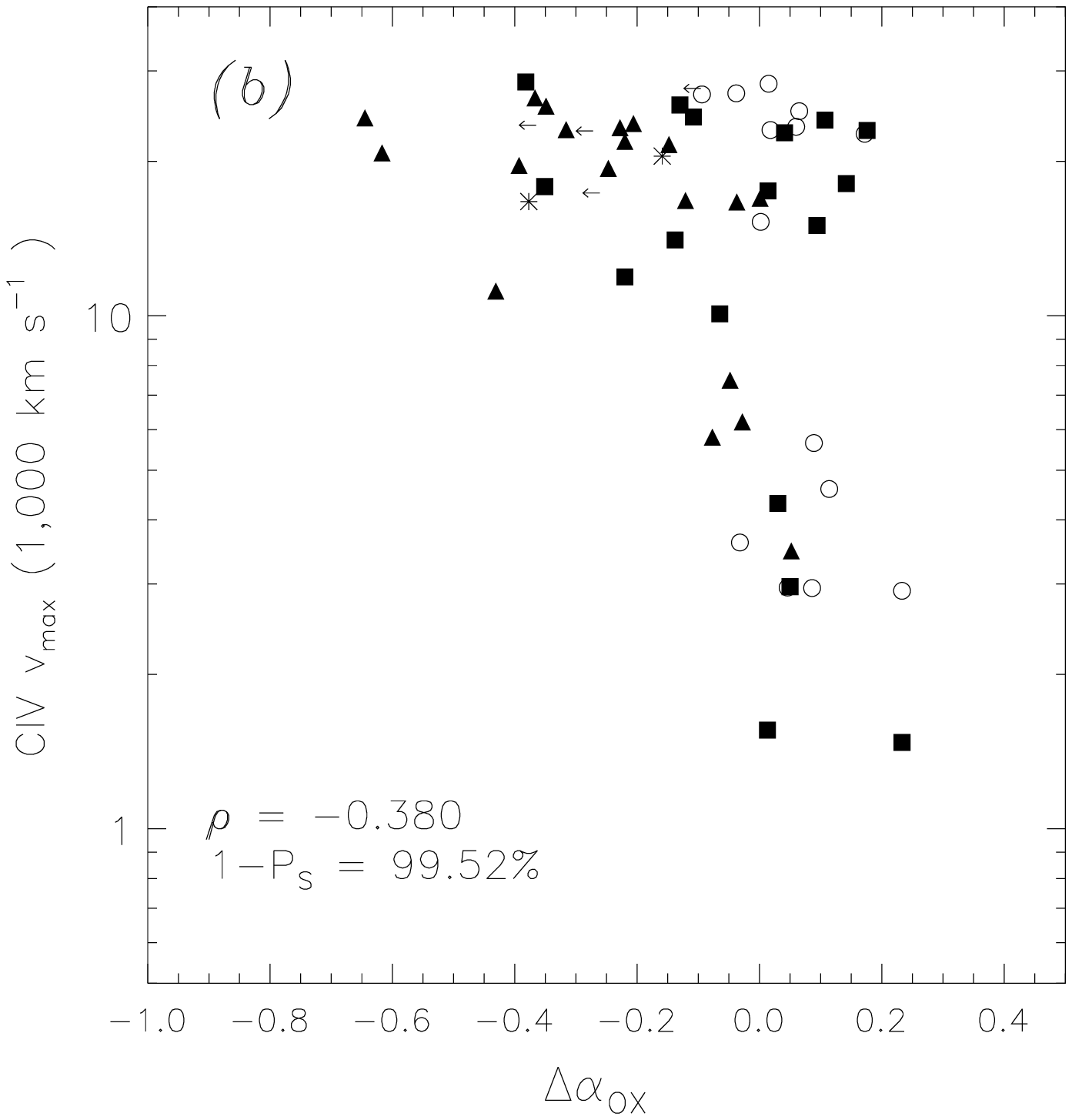}\\
    \includegraphics[width=2.4in]{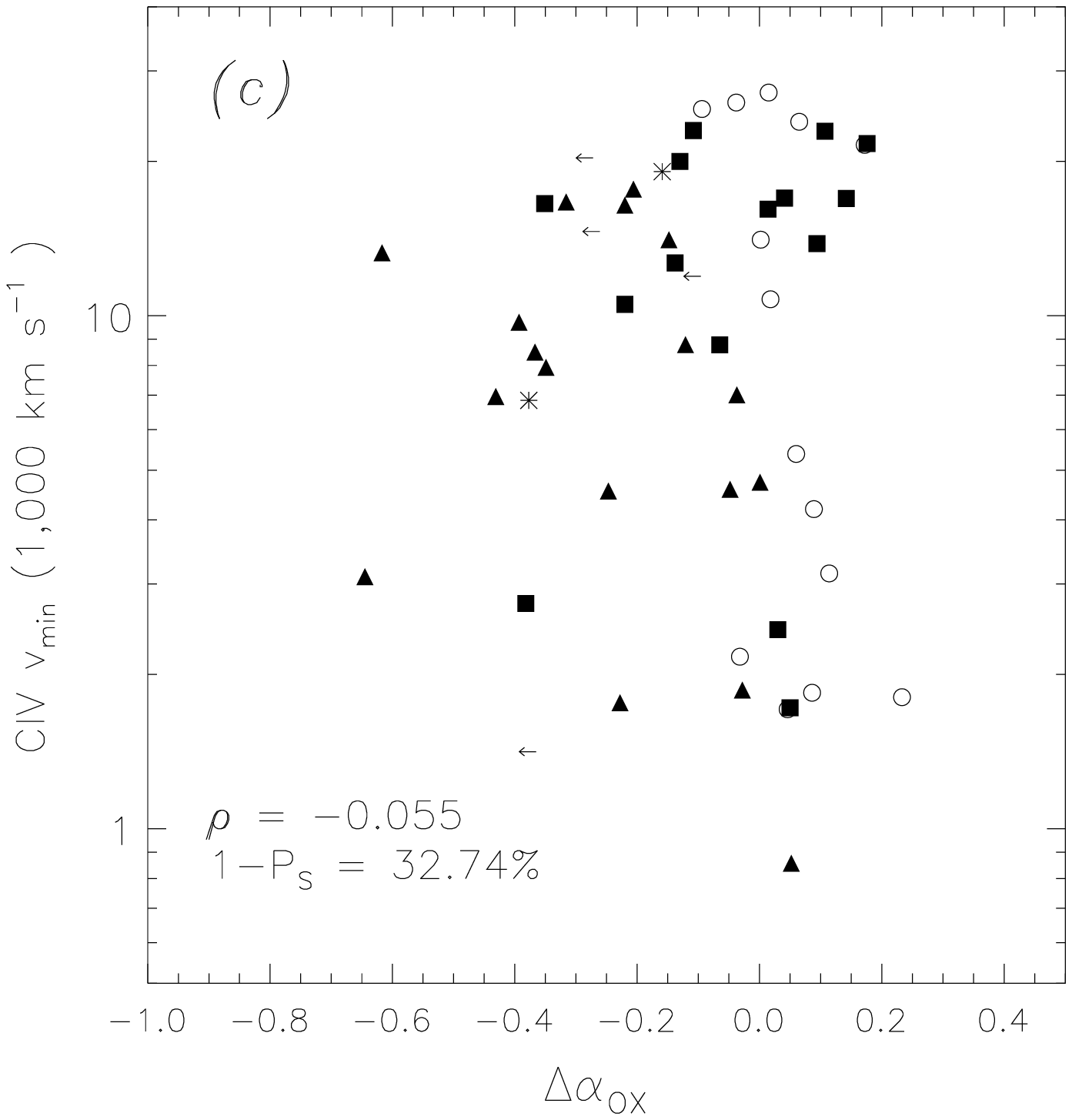}
    \includegraphics[width=2.4in]{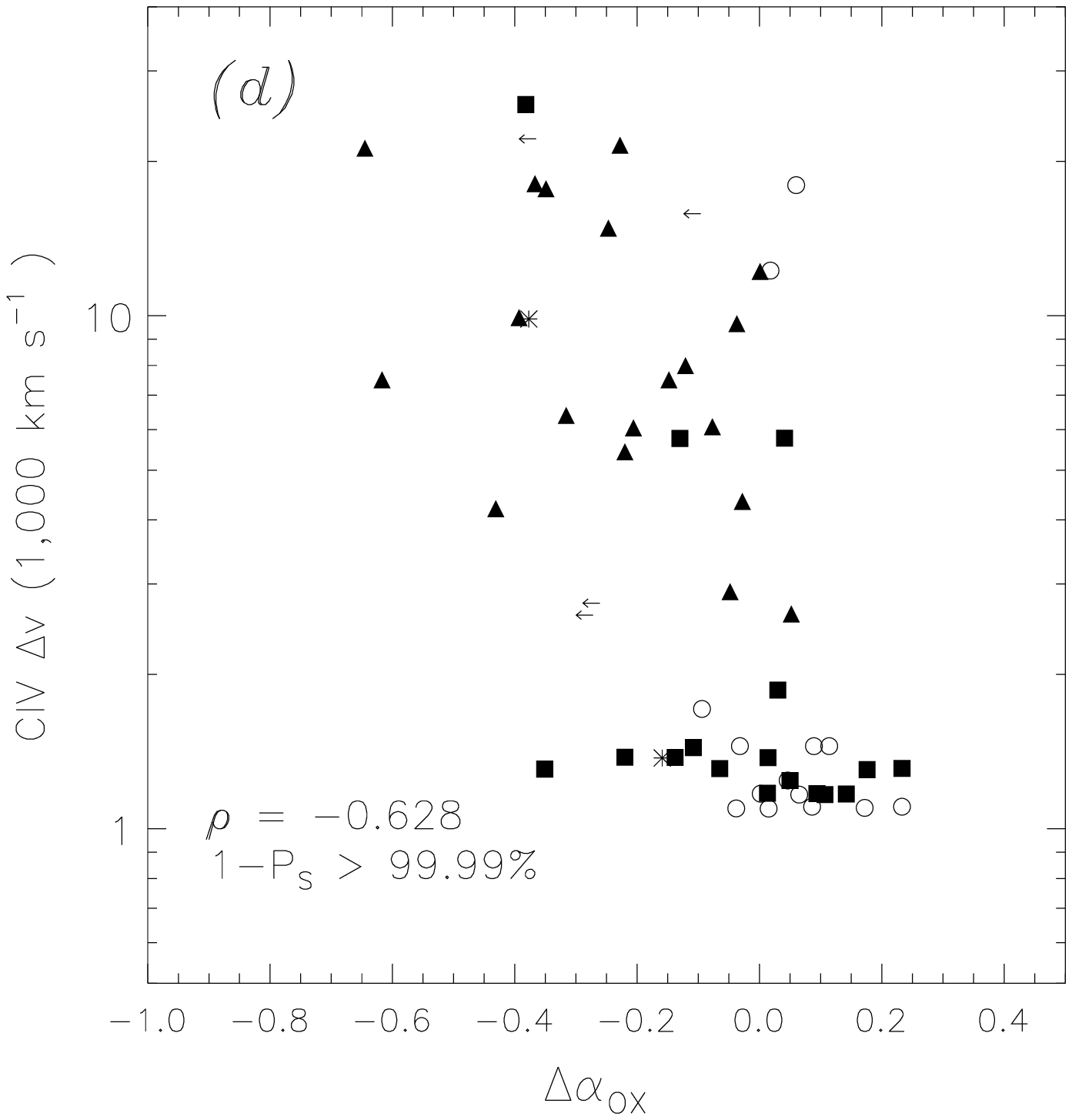}\\
    \includegraphics[width=2.4in]{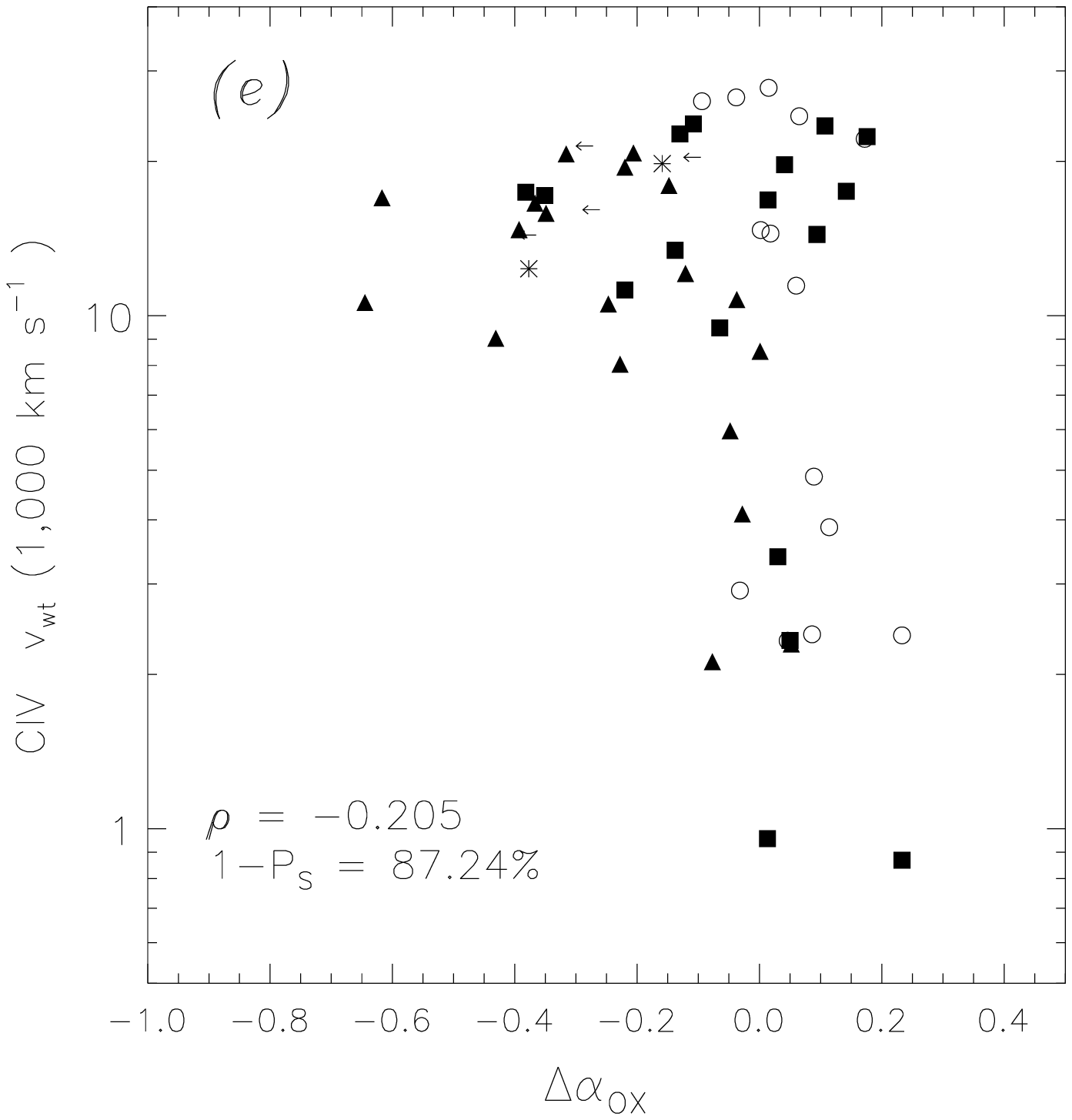}
    \caption{\footnotesize{{\it (a)}: The C~{\sc iv}~$\lambda$1549 absorption index, $AI$, plotted 
             with respect to $\Delta\alpha_{\rm ox}$ for our combined 
             {\it Chandra} snapshot survey and archival {\it XMM-Newton} 
             radio-quiet mini-BAL quasar sample (open circles), as well as the 
             detected \hbox{radio-quiet} HiBAL (filled triangles) and mini-BAL 
             (filled squares) quasars from G09. Upper limits 
             for the G09 undetected HiBAL quasars are 
             indicated by arrows, while the undetected mini-BAL quasar upper 
             limits are indicated by asterisks. All of the objects have $SN_{1700}\;>\;9$. 
             {\it (b)}: The C~{\sc iv}~$\lambda$1549 maximum outflow velocity, 
             $v_{\rm max}$, plotted versus $\Delta\alpha_{\rm ox}$. 
             {\it (c)}: The C~{\sc iv}~$\lambda$1549 minimum outflow velocity, 
             $v_{\rm min}$, plotted versus $\Delta\alpha_{\rm ox}$. 
             {\it (d)}: The C~{\sc iv}~$\lambda$1549 velocity difference,
             $\Delta v\;=\;|v_{\rm max} - v_{\rm min}|$ plotted versus 
             $\Delta\alpha_{\rm ox}$.
             {\it (e)}: The C~{\sc iv}~$\lambda$1549 weighted average velocity
             $v_{\rm wt}$, plotted versus $\Delta\alpha_{\rm ox}$.
             The numbers in the parentheses of the $y$-axis titles ($1,000$~km~s$^{-1}$)
             are the units for corresponding quantities. The results of the Spearman rank-order correlation
             analyses (correlation coefficient $\rho$ and probability $1-P_S$; also listed in Table~6) are 
             shown in each panel. }}
             \label{fig9}
\end{figure}

\begin{figure}[htbp]
    \centering
    \includegraphics[width=2.6in]{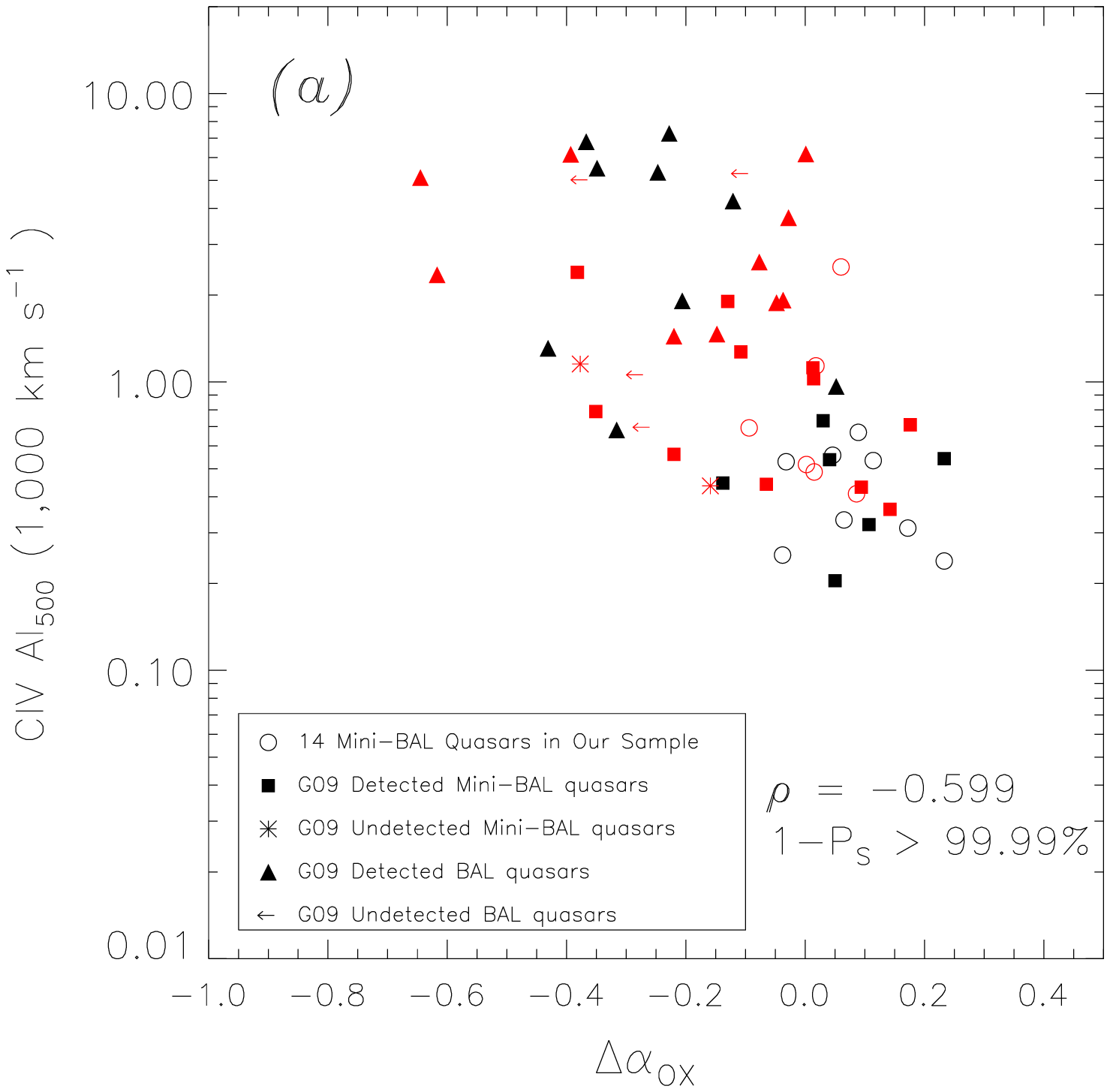}
    \includegraphics[width=2.6in]{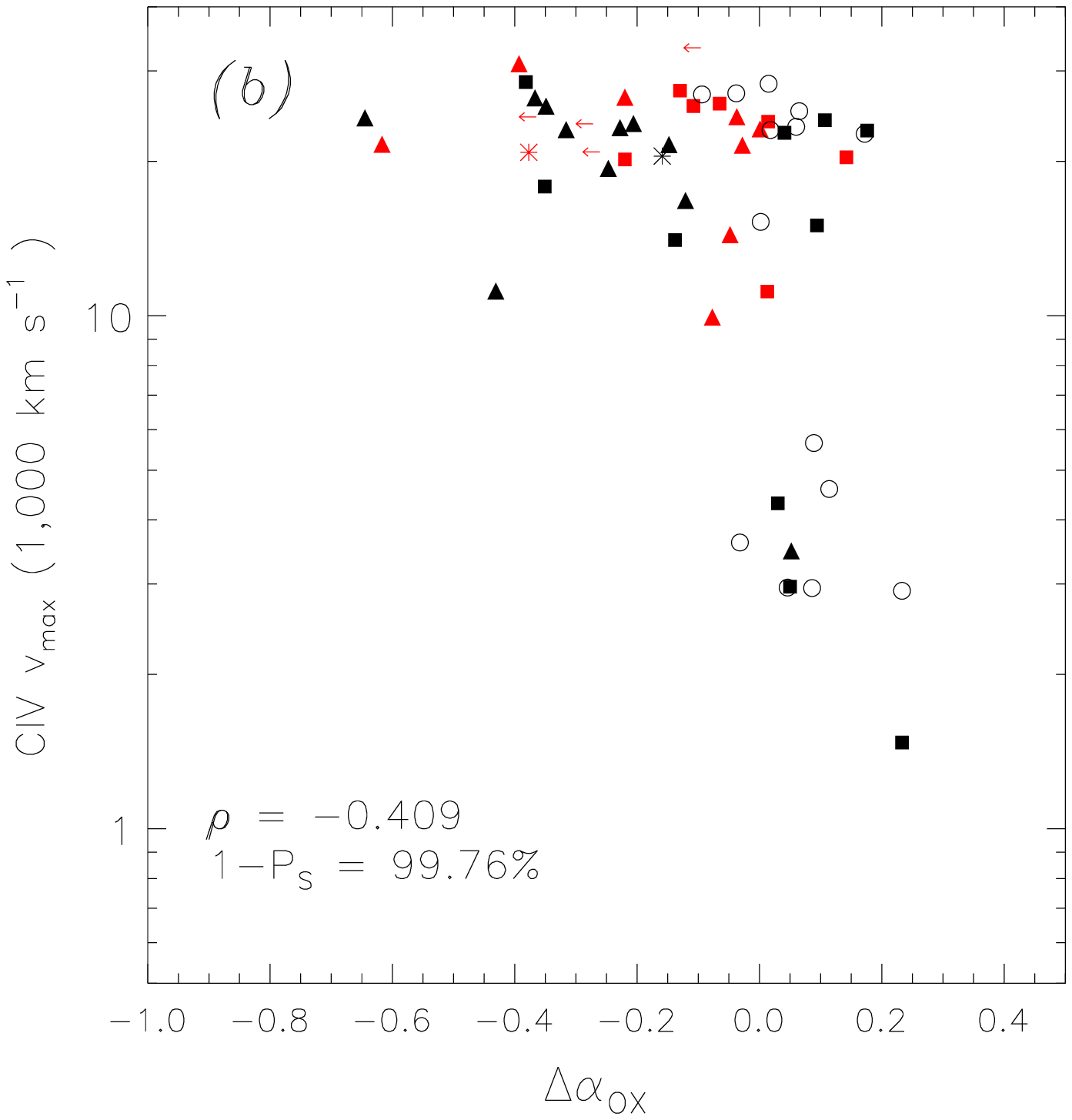}\\
    \includegraphics[width=2.6in]{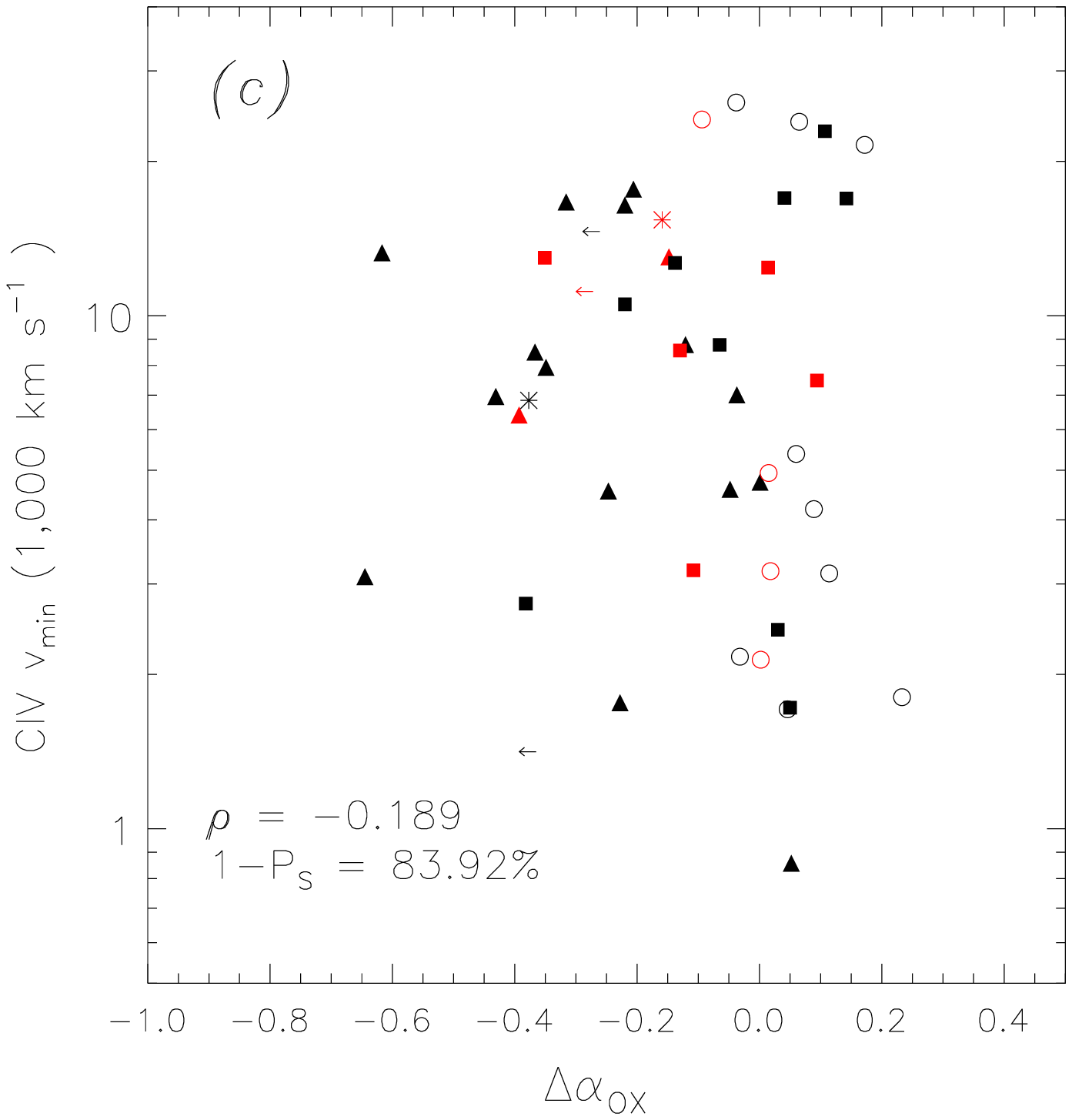}
    \includegraphics[width=2.6in]{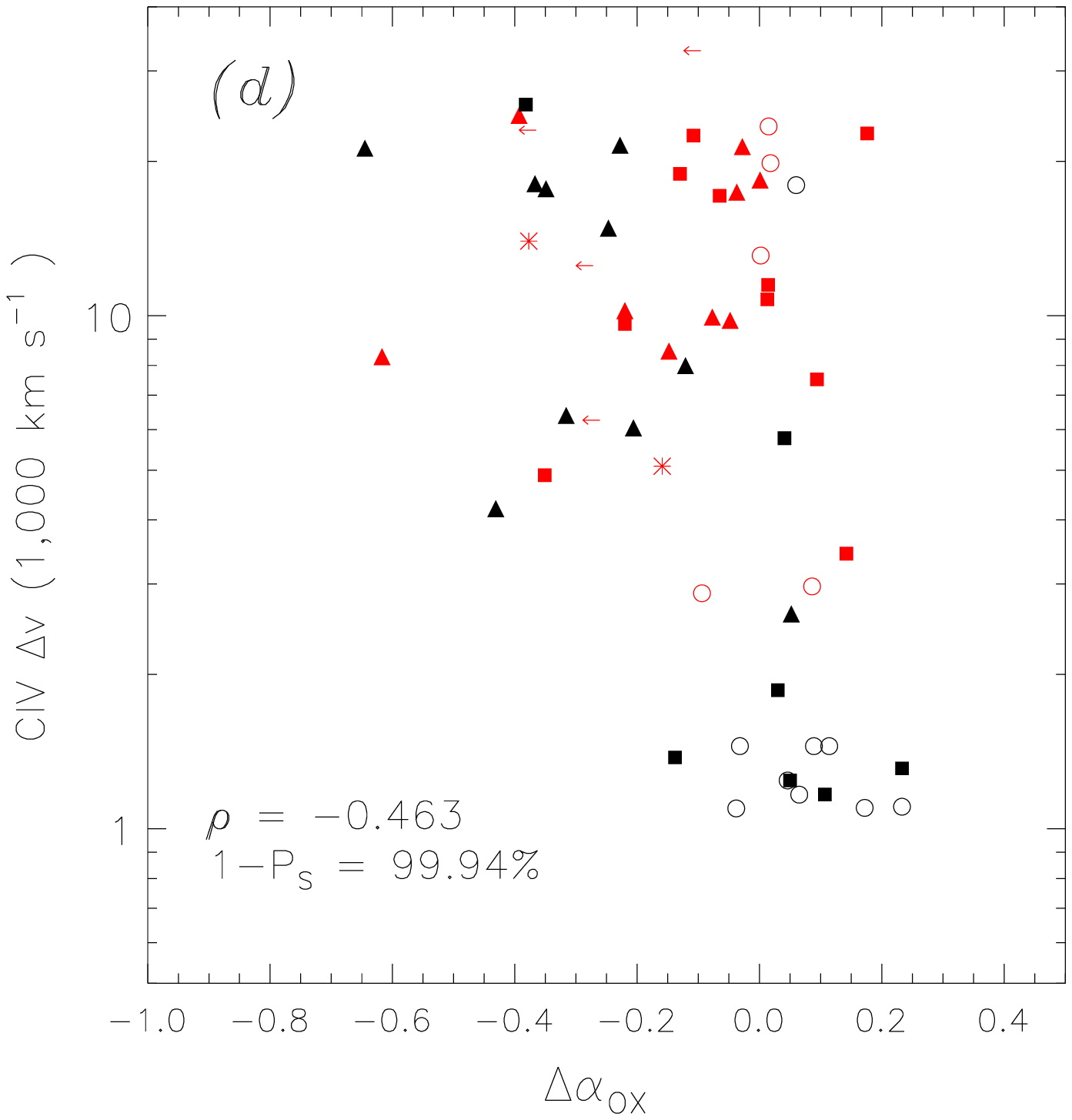}\\
    \includegraphics[width=2.6in]{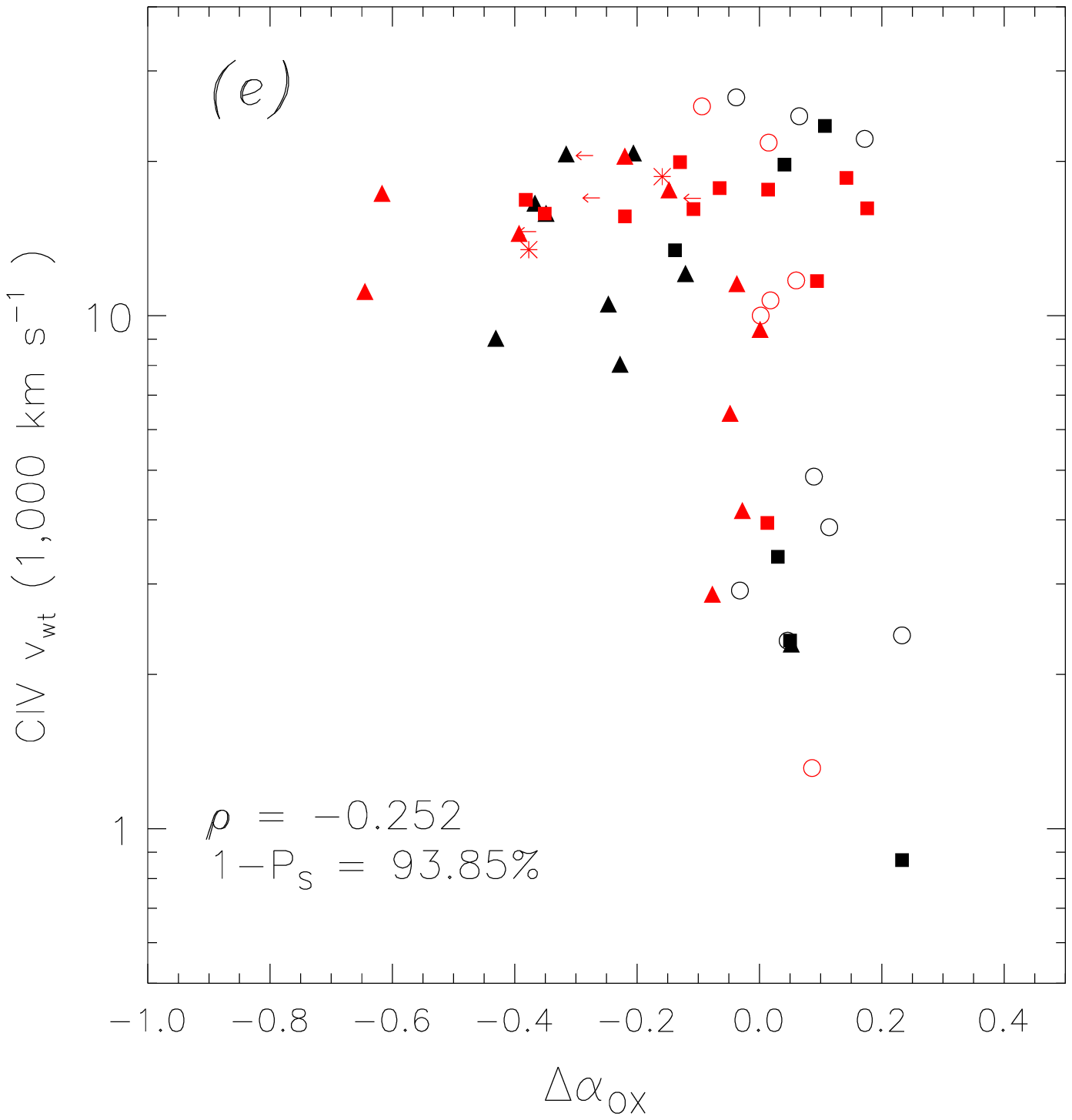}
    \caption{\footnotesize{The UV absorption properties, $AI_{500}\ (a)$, $v_{\rm max}\ (b)$, $v_{\rm min}\ (c)$,
              $\Delta v\ (d)$, and $v_{\rm wt}\ (e)$ under the velocity width limit 
             of $500$~km~s$^{-1}$, plotted
             against $\Delta\alpha_{\rm ox}$. The symbols follow the same definition as in
             Fig.~9. The red symbols show the quantities that changed under the new 
             velocity width limit. The numbers in the parentheses of the $y$-axis titles 
             ($1,000$~km~s$^{-1}$) are the units for corresponding quantities. 
             The results of the Spearman rank-order correlation
             analyses (correlation coefficient $\rho$ and probability $1-P_S$; also listed in Table~6) are 
             shown in each panel.}}
             \label{fig10}
\end{figure}

\begin{figure}[htbp]
    \centering
    \includegraphics[width=3.0in]{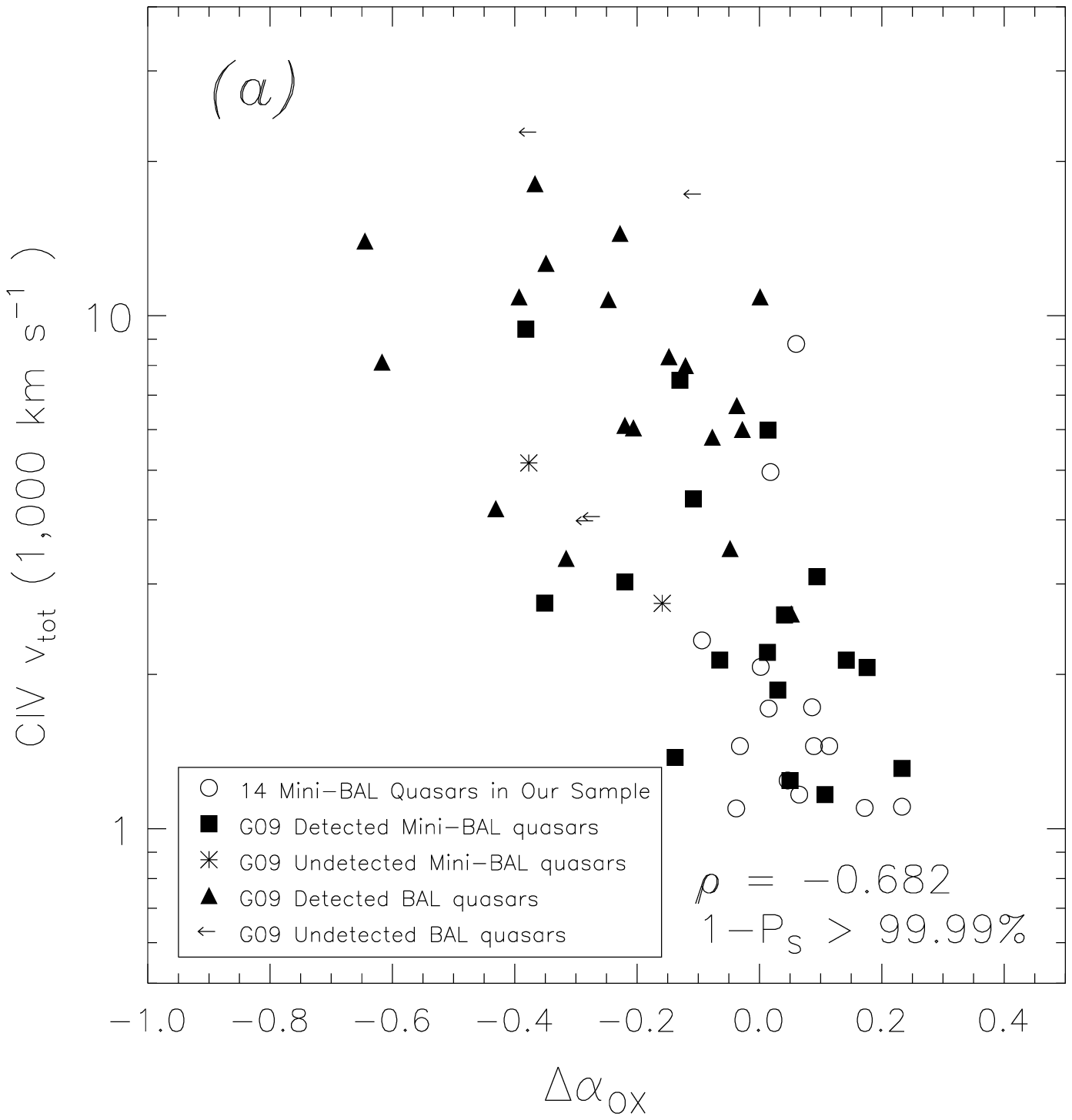}
    \includegraphics[width=3.0in]{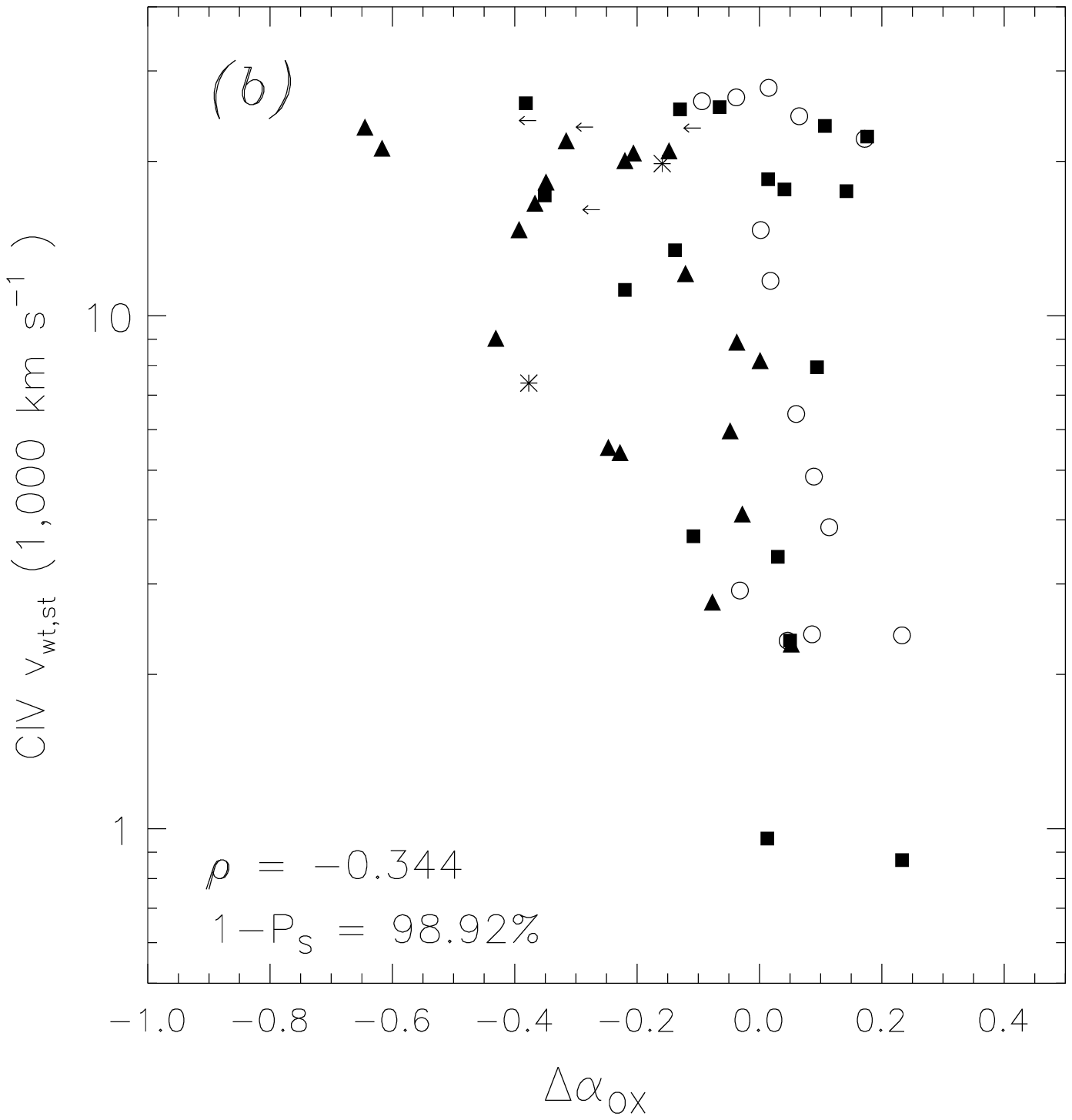}\\
    \includegraphics[width=3.0in]{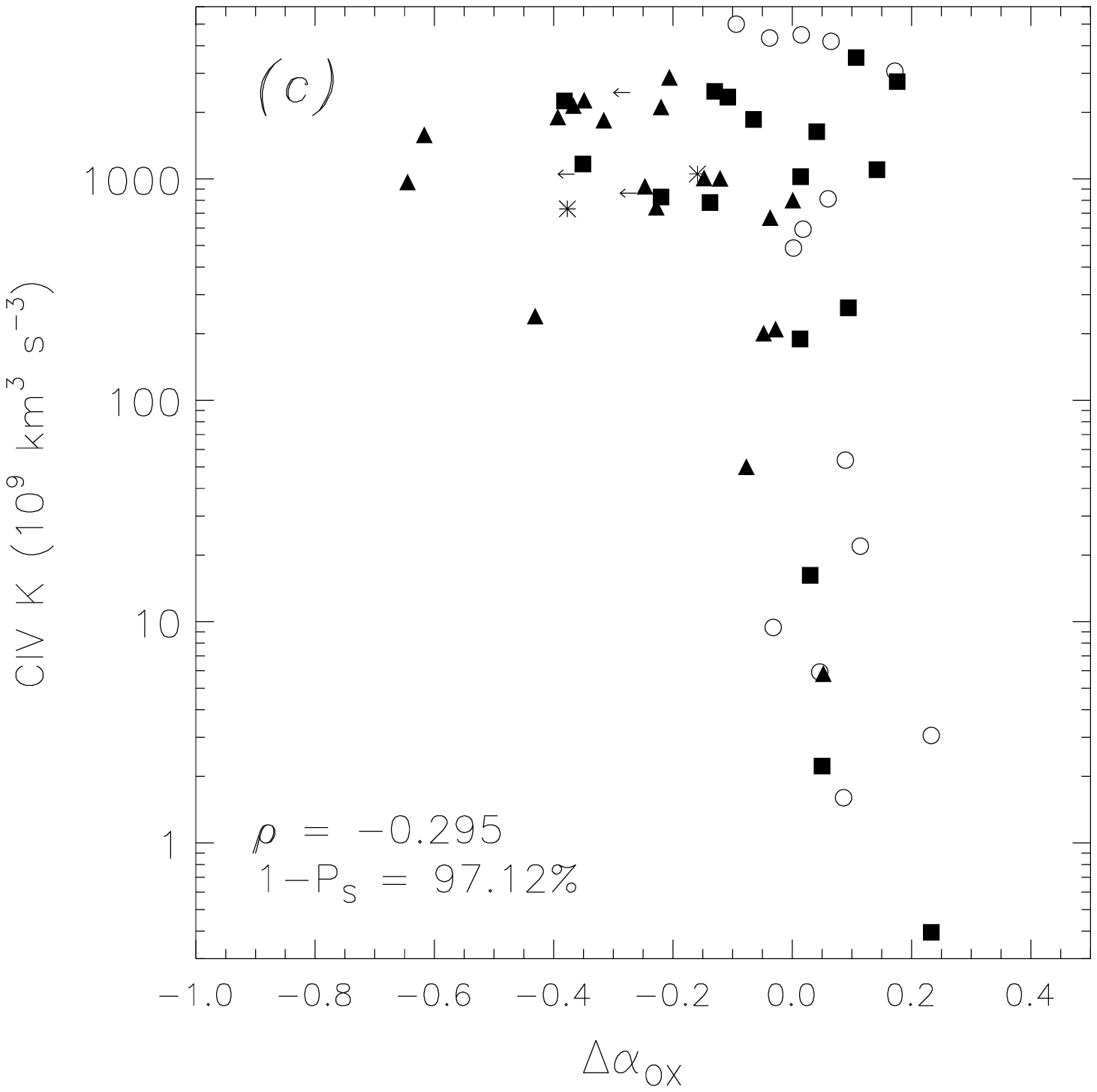}
    \caption{\footnotesize{{\it (a)}: The C~{\sc iv}~$\lambda$1549 total velocity span, $v_{\rm tot}$,
              plotted versus $\Delta\alpha_{\rm ox}$.
             {\it (b)}: The C~{\sc iv}~$\lambda$1549 weighted average velocity
             for the strongest absorption trough $v_{\rm wt,st}$, plotted versus $\Delta\alpha_{\rm ox}$.
             {\it (c)}: The C~{\sc iv}~$\lambda$1549 $K$ parameter 
              plotted versus $\Delta\alpha_{\rm ox}$.
              The symbols follow the same definition as in Fig.~9. 
              A velocity width limit of $500$~km~s$^{-1}$ is used when calculating
              $v_{\rm tot}$ and $K$ values in this figure. The numbers in the parentheses 
              of the $y$-axis titles ($1,000$~km~s$^{-1}$ or $10^9$~km$^{-3}$~s$^{-3}$)
              are the units for corresponding quantities. The results of the Spearman rank-order correlation
              analyses (correlation coefficient $\rho$ and probability $1-P_S$) are 
              shown in each panel.}}
             \label{fig11}
\end{figure}

\end{document}